\documentclass[draft]{agujournal2019}
\usepackage{url} 
\usepackage{lineno}
\usepackage{soul}
\usepackage{natbib}
\usepackage{booktabs}
\usepackage{longtable}

\newcommand{\apjs}{ApJ Suplements}
\newcommand{\aap}{A$\&$A}

\newcommand{\apjl}{ApJ Letters}

\usepackage[normalem]{ulem}
\draftfalse

\journalname{JGR: Planets}

\begin{document}

%
%

\title{Triton Haze Analogues: the Role of Carbon Monoxide in Haze Formation}

%
%



\authors{Sarah E. Moran\affil{1,2}, Sarah M. H{\"o}rst\affil{1,3,4}, Chao He\affil{1}, Michael J. Radke\affil{1},  Joshua A. Sebree\affil{5}, Noam R. Izenberg\affil{6},  V{\'e}ronique Vuitton\affil{7}, Laur{\`e}ne Flandinet\affil{7}, Fran{\c{c}}ois-R{\'e}gis Orthous-Daunay\affil{7}, C{\'e}dric Wolters\affil{7}}

\affiliation{1}{Department of Earth and Planetary Sciences,
 Johns Hopkins University,
Baltimore, MD 21218, USA}
\affiliation{2}{Bay Area Environmental Research Institute/NASA Ames Research Center, Moffett Field, CA 94035, USA}
\affiliation{3}{Hopkins Extreme Materials Institute, Johns Hopkins University, Baltimore, MD 21218, USA}
\affiliation{4}{Space Telescope Science Institute,
Baltimore, MD 21218, USA}
\affiliation{5}{Department of Chemistry and Biochemistry,
University of Northern Iowa, Cedar Falls, IA 50614, USA}
\affiliation{6}{Applied Physics Laboratory,
Johns Hopkins University, 
Laurel, MD 20723, USA}
\affiliation{7}{Univ. Grenoble Alpes, CNRS, CNES, IPAG, 38000 Grenoble, France}





\correspondingauthor{Sarah E. Moran}{smoran14@jhu.edu}


\begin{keypoints}
\item Multiple solar system bodies have complex photochemical hazes which derive from their nitrogen and carbon-rich atmospheres.
\item We generate and measure the properties of analogue hazes (``tholin'') specific to Triton-like composition and temperature.
\item Despite other similarities, Triton tholin are much more strongly oxygenated and slightly more nitrogenated than Titan and Pluto tholin.
\end{keypoints}
%
%

%
%


\begin{abstract}
Triton is the largest moon of the Neptune system and possesses a thin nitrogen atmosphere with trace amounts of carbon monoxide and methane, making it of similar composition to that of the dwarf planet Pluto. Like Pluto and Saturn's moon Titan, Triton has a haze layer thought to be composed of organics formed through photochemistry. Here, we perform atmospheric chamber experiments of 0.5\% CO and 0.2\% CH$_4$ in N$_2$ at 90 K and 1 mbar to generate Triton haze analogues. We then characterize the physical and chemical properties of these particles. We measure their production rate, their bulk composition with combustion analysis, their molecular composition with very high resolution mass spectrometry, and their transmission and reflectance from the optical to the near-infrared with Fourier Transform Infrared (FTIR) Spectroscopy. We compare these properties to existing measurements of Triton's tenuous atmosphere and surface, as well as contextualize these results in view of all the small, hazy, nitrogen-rich worlds of our solar system. We find that carbon monoxide present at greater mixing ratios than methane in the atmosphere can lead to significantly oxygen- and nitrogen-rich haze materials. These Triton haze analogues have clear observable signatures in their near-infrared spectra, which may help us differentiate the mechanisms behind haze formation processes across diverse solar system bodies.
\end{abstract}
\section*{Plain Language Summary}
Triton is the largest moon of the outer planet Neptune. It has a very thin atmosphere made of similar gases to the atmospheres of the dwarf planet Pluto and Saturn's moon Titan. Sunlight or high energy particles can break apart the molecules that make up these gases, which can then react to form solid particles, called hazes. We made haze particles in an atmospheric chamber under Triton-like temperature (90 K) and atmospheric composition (small amounts of carbon monoxide and methane in molecular nitrogen), and then measured the chemical and physical properties of the resulting material. We compare our results to similar measurements of laboratory materials made for Pluto and Titan. Our results show larger oxygen and nitrogen contents for these Triton particles, suggesting that increasing carbon monoxide in the atmosphere changes the chemistry of hazes. Within the laboratory hazes, we see signatures of molecular bonds containing oxygen in the near-infrared, which might be useful for identifying these species with future observations of or missions to Triton.
%
%

\section{Introduction}
%

%
%
%
%

Triton, as Neptune's largest moon, is unique among the ice giant moons because of its thin nitrogen atmosphere and its status as a captured Kuiper Belt Object (KBO) \citep{mckinnon1984,agnor2006}. Also considered a candidate ocean world, Triton is a natural comparison to two other worlds, Titan and Pluto, which also have nitrogen-rich atmospheres with trace amounts of carbon monoxide and methane, though the absolute mixing ratios differ between the three. Rich photochemistry has been observed both on Titan and Pluto from dedicated spacecraft observations by the Cassini-Huygens mission to the Saturn system and the New Horizons flyby of Pluto. Dramatic haze layers are seen in the atmospheres of both worlds from these two missions \citep{Porco2005,stern2015}. Voyager 2 observations also suggest that Triton has a haze layer and potentially nitrogen ice clouds \citep{herbert1991,ragesandpollack1992,strobel1995tritonatmosandiono,yelle1995tritonatms}, though the haze properties, especially as compared to that of Titan and Pluto, remain poorly characterized due to Voyager 2 phase angle and detection limit constraints \citep{pollack1990}.

Triton's atmosphere, due to its extremely cold surface temperature of 38 K \citep{conrath1989}, is in vapor pressure equilibrium with its surface ices, making atmosphere-surface interactions active and strongly seasonally dependent \citep{hansen1992n2seasonal,cruikshank1993n2surface}. Similar activity is observed for Pluto \citep{lewis2019plutoseasons}, where seasonal sublimation appears to drive winds \citep{tefler2018dunespluto}. Measurements of Triton's atmospheric pressure at the surface range from 14$\pm$1 $\mu$bar in 1989 from Voyager 2 radio science \citep{tyler1989} to 19$\pm$1.8 $\mu$bar from stellar occultations in 1995 and 1997 \citep{olkin1997,Elliot1998}, suggestive of seasonal sublimation and deposition. Nitrogen dominates the atmosphere, and vertical profiles of the N$_2$ \citep{krasnopolsky1993} and CH$_4$ content, which decreases with altitude \citep{herbert1991}, have been known since Voyager 2. From the surface at 38 K, the upper atmosphere reaches temperatures of approximately 90 K \citep{strobelandzhu2017}. With the Very Large Telescope of the European Southern Observatory, \citet{lellouch2010} was able to measure the amount of CO in the atmosphere, where only upper limits were achieved by Voyager 2. \citet{lellouch2010} found a partial pressure of 24$\pm$3 nbar for CO, but did not measure a surface pressure at the time of observations.

Triton's young ($\leq$ 10 Ma) surface \citep{sternandmckinnon2000tritonsurfaceage,schenkandzahnle2007surfaceage} requires geological activity suggestive of significant internal heating, potentially provided by obliquity tides \citep{nimmoandspencer2015}, which could maintain a subsurface liquid ocean. Plumes/geysers on Triton, observed by Voyager 2, also suggest seasonally driven winds and/or subsurface activity \citep{Hansen1990windsplumes,hammond2018}. Organic material generated by photochemistry in the atmosphere could then interact with this potential subsurface ocean, or even the plume outflow material itself, furthering prebiotic chemical reactions, as may also occur on Titan \citep{neish2010}. 

In Titan's atmosphere, the dissociation of N$_2$, CH$_4$, and other minor species that initiate haze formation is primarily driven by solar UV radiation \citep{vuitton2019}, while  galactic cosmic ray impacts can also initiate ionization reactions to a lesser degree \citep{Gronoff2009ionizationmagfieldtitan,gronoff2011}. Magnetospheric ions from Saturn on the order of up to 1 keV with flux 10$^{-6}$ cm$^{-2}$ s$^{-2}$ have been detected at Titan \citep{hartle2006} and are a minor driver of atmospheric chemistry on the moon. However, at Triton, Neptune's magnetospheric particles interact with Triton's ionosphere with energies an order of magnitude higher \citep{krimigis1989,stone1989,thompson1989tholin} and thus may play an increased role in atmospheric chemistry, though solar EUV photons likely still dominate \citep{lyons1992}. The flux of solar photons naturally decreases with distance from the Sun from Titan to Triton to Pluto, from $\sim$15 W m$^{-2}$ to $\sim$1.5 W m$^{-2}$ to $\sim$0.9 W m$^{-2}$. Determining how far organic atmospheric chemistry has proceeded on Triton as a result of haze formation processes, and how this compares to Pluto and Titan, is thus of high interest given the similarities in bulk composition yet major differences in energy distribution between the three planetary bodies.

Efforts to model the haze formation process on Pluto and Triton have shown previously that the CO and CH$_4$ mixing ratios in the atmosphere act as a strong control on the overall composition of haze and photochemical gas products \citep{krasnopolsky1995tritonphotochem,krasnopolsky2012,strobelandzhu2017}. These models have noted that ethylene gas, C$_2$H$_4$, as a product of this photochemical process is likely to condense the most readily of all photochemical products considered \citep{krasnopolsky1992,wong2017}. Hazes of the upper atmosphere can act as thermal controls of both the atmosphere and surface \citep{zhangandstrobel2017}, which is critically dependent on the exact composition and optical properties of the material. Recent coupled photochemical and microphysical modeling suggests that gas phase, C$_2$-based hydrocarbon adsorption onto aerosol particle surfaces \citep{luspaykuti2017} as well as ice condensation onto haze particle condensation nuclei (dominated by HCN cores) may better explain the observations of both Pluto and Triton's atmosphere than photochemical haze alone \citep{lavvas2021}, where C$_2$H$_4$ ice would dominate the composition of these heterogeneous, coated particles on Triton \citep{krasnopolsky2020,lavvas2021}. Both spherical and fractal aggregate particles can explain Triton observations using heterogeneous haze-ice particles, though fractals are preferred \citep{ohno2021triton}.

Current models are in need of additional data to better constrain these atmospheric processes, but \textit{in situ} missions to the outer solar system to characterize atmospheric chemistry are both few and far between, and often only reveal far more complexity than previously assumed, as in the case of both Titan and Pluto. Moreover, because of the unique and highly complex chemistry occurring in these haze layers, theoretical models often cannot fully capture haze formation from initial gases to complex haze molecules due to both a lack of chemical kinetics of the relevant chemical reactions as well as computational expense \citep{berry2019,vuitton2019}. As such, laboratory synthesis of atmospheric hazes has been performed to generate and study analogue particles to those potentially made in planetary atmospheres, including those for Titan \cite[e.g.,][]{khare1984,imanaka2004,vuitton2010HCN}, the Early Earth \cite[e.g.,][]{trainer2006,dewitt2009,hasenkopf2010,horst2018earlyearth}, and even exoplanets \cite[e.g.,][]{horst2018production,gavilan2018organicaerosols}. These experiments have revealed a wealth of information about haze formation and properties, including their effect on spectra \citep{brasse2015}, the production of prebiotic molecules \citep{horst2012formation,sebree2018}, and potential chemical pathways to haze formation \citep{gautier2014orbitrap,gautier2016hplcorbitrap}.

Previous atmospheric haze formation experiments also exist for Triton specifically \citep{mcdonald1994titanandtritontholin,thompson1989tholin}, but these past experiments included only nitrogen and methane in their starting gas mixtures. Thanks to more recent ground-based observations \citep{lellouch2010}, we now know that CO is in fact the second most abundant molecule in Triton's N$_2$ atmosphere, unlike the Plutonian and Titanian atmospheres where CH$_4$ is found in higher mixing ratios than CO \citep{krasnopolsky2012}. Some experiments have included CO in Titan-like mixtures \citep{bernard2003,coll2003,Tran2008}, suggesting oxygenated molecules play a role in the gas phase chemistry. More recent laboratory experiments have shown that an initial inventory of gas phase CO also substantially affects resulting tholin particles \citep{horstandtolbert2014,he2017co}. CO was included in recent atmospheric experiments of a Pluto-like atmosphere \citep{JovanovicPlutoOrbitrap, jovanovic2021opticalconstants}, with substantial impact on the tholin produced. However, this Pluto experiment, as well as the previous Triton experiments, were performed at room temperature, though temperature should also influence the chemistry occurring in the atmosphere. Moreover, the published spectra and chemical measurements from every N$_2$-CH$_4$-CO experiment to date has been performed with mixing ratios where CH$_4$ $\geq$ CO, as is more representative of Titan and Pluto rather than Triton. The only published experiments performed with CO $>$ CH$_4$ only examined the particle density and size of the resulting solid \citep{horstandtolbert2014}, not its chemistry. Therefore, in this study, we perform a new set of atmospheric chamber experiments for a Triton-like atmosphere -- including this second most abundant molecule of CO at a higher percentage than CH$_4$ and at Triton-like temperature -- to examine the chemical and physical properties of hazes that may influence Triton's atmosphere and climate, surface, and planetary evolution. 

\section{Methods} \label{sec:method}
First, we generated Triton haze analogue particles within the PHAZER (Planetary HAZE Research) chamber \citep{he2017co} under Triton-relevant temperature and pressure with an AC cold plasma discharge as the energy source. We subjected the resulting solid particles to combustion analysis to obtain the elemental ratios of C, H, O, and N that make up the particles. Next, we performed a more in-depth analysis of the chemical composition of the solids produced in the PHAZER chamber by utilizing very high resolution Thermo Fisher Scientific LTQ-Orbitrap XL mass spectrometry. Once mass spectrometer measurements were complete, we employed custom software called \textit{idmol} \citep{horstthesis} to make molecular identifications and perform data analysis from the mass spectrometry data. Finally, we obtained the transmittance and reflectance of the Triton haze analogues by measuring thin films of the particles with a Bruker Vertex 70V Fourier Transform Infrared Spectrometer. 

\subsection{Triton Haze Analogue Production}

\begin{figure*}[hbt!]
\centering
\includegraphics[angle=0,width=0.95\linewidth]{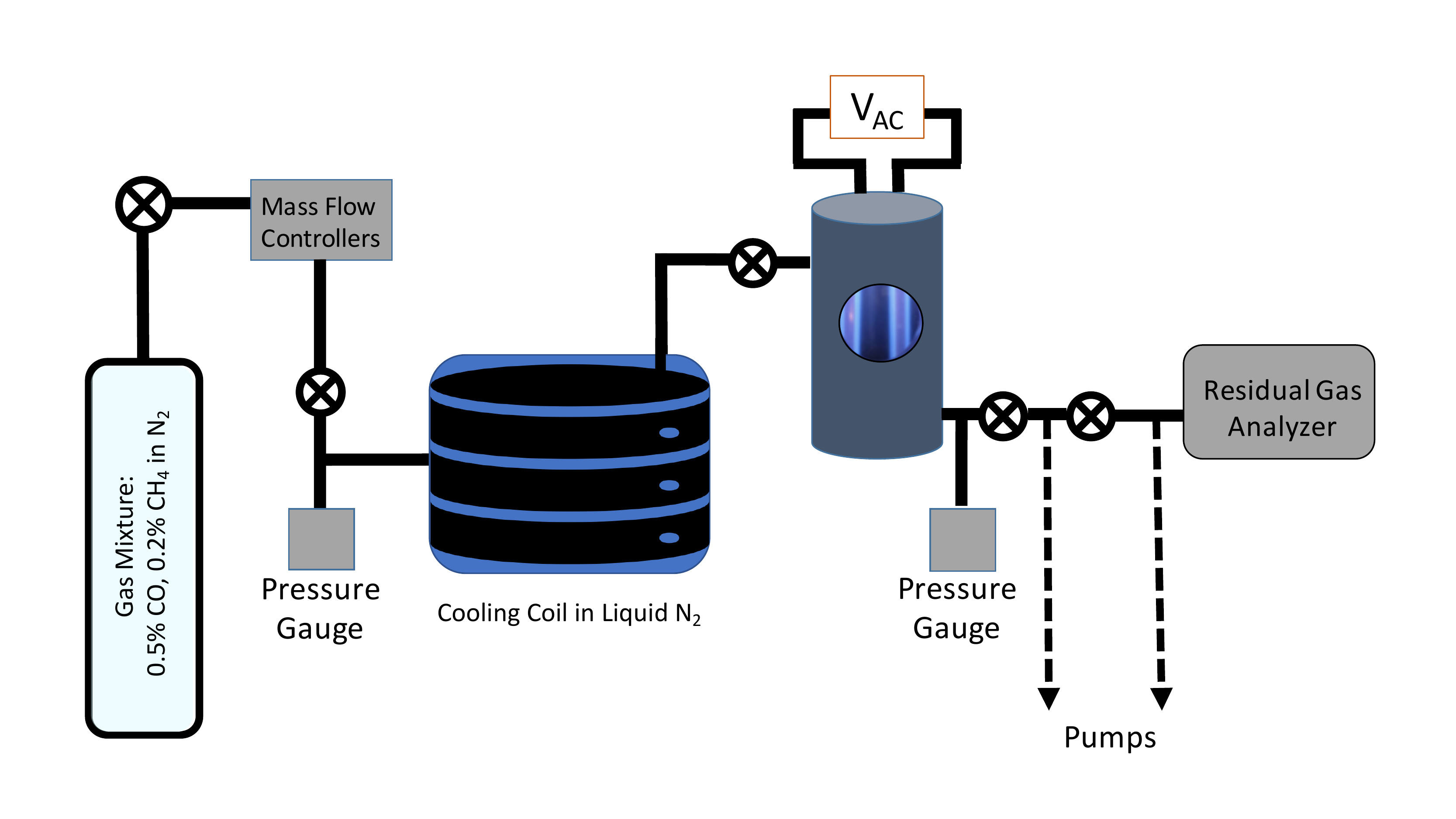}
\caption{Schematic of the PHAZER chamber set-up at Johns Hopkins University. The 0.5\% CO, 0.2\% CH$_4$ in N$_2$ gas mix is flowed through the cooling coil submerged in liquid nitrogen before flowing into the reaction chamber. The mixing ratios were chosen to represent the high end of atmospheric CO from existing observations \cite[e.g.,][]{krasnopolsky2012}. The mass flow controllers are set at 10 sccm so that the pressure in the chamber is 1 mbar. The gas mixture is then exposed to the AC glow discharge, and analogue haze chemistry proceeds at 90 K.}
\label{fig:phazer}
\end{figure*}

With the PHAZER chamber and accompanying apparatus (JHU, Baltimore, MD), we produced Triton haze analogue particles. We continuously flowed a gas mixture of 0.5\% CO (Airgas) and 0.2\% CH$_4$ (Airgas) in N$_2$ (Airgas) for 72 hours, as is standard PHAZER protocol \citep{he2017co} for the generation of substantial macroscopic sample. To calculate these mixing ratios, we use the partial pressure of CO determined by \citet{lellouch2010}, 24 nbar within a factor 3. We divide the upper limit of this partial pressure by the surface pressure determined by Voyager 2, 14 $\mu$bar, to obtain 0.5\% CO in our mixture. \citet{lellouch2010} reports the surface partial pressure ratio CO/CH$_4$ to be $\sim$2.5, resulting in our CH$_4$ mixing ratio of 0.2\%. While our experimental ratio of CO to CH$_4$ reflects Triton's atmosphere, their absolute values compared to N$_2$ are higher than the atmospheric mixing ratios reported by \citet{lellouch2010} since we include this surface pressure correction. These values are chosen to generate higher yields of tholin in a reasonable experimental timeframe. We discuss the implications of this approximation on our results in Section \ref{sec:icecondensation}.

The gas mixture first is flowed through the cooling coil immersed in a liquid 77 K N$_2$ bath, and then into the reaction chamber so that gases in the chamber are approximately 90 $\pm$5K \citep{he2017co}, following the best fit upper atmospheric temperature determined by \citet{strobelandzhu2017}. This temperature corresponds to regions 300 km above Triton's surface where Triton's ionosphere is located \citep{tyler1989}. On Titan, haze formation begins in the ionosphere \citep{Liang2007,lavvas2013titanionosphere}, and Neptune's magnetosphere can drive similar reactions on Triton \citep{thompson1989tholin}. However, the main haze layer is observed much lower, from 4 km to 30 km \citep{ragesandpollack1992}, where the atmospheric temperature is closer to 50 K \citep{strobelandzhu2017}. Our laboratory set-up cannot be maintained at this low a temperature, as 90 K is near our experimental limit. Nevertheless, running at 90 K represents a much more accurate temperature condition than any previous Triton haze experiments, which have all been performed at room temperature.

An AC cold plasma glow discharge then provides an energy density of order 170 W m$^{-2}$ into the system \citep{he2019}. This flux is applied over 72 hours, for a total of 4.4$\times$10$^7$ Wm$^{-2}$ over the course of the experiment. Given that Triton receives $\sim$1.5 Wm$^{-2}$ in incident solar radiation, our experiment corresponds to approximately 340 days of solar irradiation at Triton. The plasma source is energetic enough to dissociate even extremely stable molecules, such as N$_2$ and CO \citep{cabletholinreview2012}. This cold plasma does not directly replicate any single energetic atmospheric process, photochemical or otherwise. Instead it provides a method by which to approximate the energetics of upper atmospheres. For Triton's upper atmosphere, such energy distributions may derive from a combination of sources such as incident UV solar photons, cosmic ray bombardment, or charged particles from the magnetic field of Neptune. 

The gases were flowed at 10 standard cubic centimeters per minute (sccm) so that the pressure within the chamber was maintained at 1 mbar, which is higher than any pressure in Triton's atmosphere, which at highest is estimated to be 40 $\mu$bar \citep{lellouch2010}. Maintaining the experimental chamber at 1 mbar pressure reduces reaction times given the mean free path within the size of the reaction chamber, allowing for completion of experimental runs within a reasonable timeframe.
Furthermore, many other haze formation experiments are run at this pressure \cite[e.g.,][]{khare1984,he2017co,JovanovicPlutoOrbitrap}, which allows for comparison of the chemical and physical properties of the resulting tholins as a function only of gas species and temperature without the additional variable of pressure, which previous work has shown to affect the tholin composition and spectra. For example, \citet{imanaka2004} found that tholin produced at lower pressures (0.13 mbar) incorporated more nitrogen than tholin produced at higher pressure (23 mbar), and that this led to stronger, deeper features from N-H bonds and aromatic ring structures in their spectra.

As ionization and dissociation of the inital N$_2$, CO, and CH$_4$ gas molecules proceeds, these ions and radicals react to make new molecules, and some combination of initial and newly created molecules generate longer and longer molecular chains. Eventually, some such molecules become macroscopic solids and deposit out in the form of orange-brown powder on the chamber walls, floor, and quartz disks placed at the bottom of the chamber. We use quartz substrates because quartz is inert and does not react with either the energy source or gas mixture during the experiment, and is well characterized in our reaction chamber \citep{He2021titanftir}.

After 72 hours, the gas flow is turned off and remains under vacuum for 48 hours to allow volatile dissipation. The chamber is then moved into a dry ($<$ 0.1 ppm H$_2$O), oxygen-free ($<$ 0.1 ppm O$_2$) N$_2$ glove box (Inert Technology Inc., I-lab 2GB) where solid sample is collected and stored, insulated from ambient atmosphere and light sources. To determine the production rate of the experiment, we manually collect solid material from the chamber walls and base and weigh the total with an analytical balance (Sartorius Entris 224-1S with standard deviation of 0.1 mg). The production rate is then calculated by dividing the total mass by the runtime of the experiment (72 hours), with the error (0.20 mg/hr) induced from the balance and the sample residuals on the reaction chamber walls. We note also that the production rate is necessarily a lower limit as collection efficiency is always less than 1. Samples were stored in the glovebox for $\sim$6 months prior to mass spectrometry measurements, $\sim$21 months prior to spectroscopy measurements, and $\sim$30 months prior to combustion measurements. Agreement between combustion analysis results and Orbitrap mass spectrometry results, discussed in Section \ref{sec:compositionresults}, confirms that external chemical alteration does not take place in storage in the glovebox.

\subsection{Combustion Analysis}
We employed elemental combustion analysis with a Thermo Scientific Flash 2000 Elemental Analyzer (Department
of Chemistry and Biochemistry, University of Northern Iowa, IA, USA) of the Triton haze analogues produced in the PHAZER chamber. We placed 1 mg of particles in the analyzer to directly measure percentages of C, H, and N. We then perform mass subtraction to calculate the percentage of O.

\subsection{Fourier Transform Infrared Spectroscopy}
Using a Bruker Vertex 70V Fourier Transform Infrared Spectrometer (JHU, Baltimore, MD), we measured the transmission and reflectance of thin films of Triton haze analogue deposited on quartz substrates from 0.4 to 5 microns. From 0.4 to 1.1 microns, we use a Si-diode detector while from 0.83 microns to 5 microns, we employ a DLaTGS detector. Overlap between detectors allows for calibration between the wavelength ranges. The spectrometer is fitted with a quartz beamsplitter, and uses a near-IR source of a tungsten-halogen lamp. A silicon carbide globar provides the mid-IR source. When performing measurements, we vent the optical bench to $\leq$ 1 hPa. All measurements were performed at room temperature, monitored and held stable at 294 K. We first take transmission and reflectance measurements of a blank quartz disc to provide a baseline correction before measuring the Triton sample deposited on the quartz substrate. We use a source aperture size of 2 mm and average over 250 scans for each measurement. With the instrument settings as configured, our measurements have a resolution of 2 cm$^{-1}$. Interference fringes are observed in the optical to near-IR range (0.45 micron to 2 micron), and we perform a correction for this fringing following the moving average method of \citet{neri1987fringesremoval}, as in \citet{He2021titanftir}. This correction is given as 

\begin{equation}
    F(x_n)= \frac{1}{4}(2G(x_n) + G(x_{n+m}) + G(x_{n-m}))
\end{equation}

where F(x$_n$) is the fringe-removed spectrum, G(x$_n$) is the original spectrum, and 2m is the number of integer points in d, the average fringe spacing, which is 2260 cm$^{-1}$. With spectral resolution of 2 cm$^{-1}$, we find m to be 565 for our transmission and reflectance spectra. For completeness, we provide the uncorrected spectrum with fringes visible in \ref{sec:appendix_fringes}.

Once fringe removal is complete, we then compare the spectral features observed in the Triton tholin to both previous results of relevant planetary haze analogues as well as to general spectral databases in order to identify chemical bonds and functional groups.

\subsection{Orbitrap Mass Spectrometry}
To investigate the detailed chemical composition of the Triton haze analogue sample, we employed very high resolution mass spectrometry with a Thermo Fisher Scientific LTQ-Orbitrap XL mass spectrometer with an Ion Max electrospray ionization source (ESI) (IPAG, Grenoble, FR), which has resolving power (m/$\Delta$m) of at least 10$^5$ up to m/z 400 and an exact mass accuracy of $\pm$2 ppm. Prior to sample measurement, we performed mass calibration between m/z 200 and 2000 with Thermo Fisher Scientific Calmix (caffeine, MRFA peptide, and Ultramark 1621 for positive mode and sodium dodecyl sulfate, sodium taurocholate, and Ultramark 1621 in negative mode) solution. We also measured a blank solution from m/z 150-1000 of pure CH$_3$OH with the Orbitrap to account for and remove any potential contamination in sample signal from the mass calibration solution, ambient conditions, or the sample vial. To prepare the sample, we dissolved the Triton haze analogue in CH$_3$OH (methanol) at a concentration of 1 mg/mL. We then subjected the dissolved sample to sonification for 1 hour, followed by centrifugation at 10,000 rpm for 10 minutes. The soluble fraction of the Triton sample was then diluted again in CH$_3$OH at 1 mg/mL. We injected the diluted soluble fraction into the Orbitrap with electrospray ionization (ESI) and obtained overlapping mass-to-charge (m/z) bin measurements of m/z 50-300, m/z 150-450, and m/z 400-1000. The tube lens voltage was set to 50 V, 70 V, and 90 V respectively, with source voltages of 3.5 kV (positive) and 3.8 kV (negative). For each mass-to-charge bin, our measurements averaged together four scans of 128 microscans, with the injection of sample set at a flow rate of 3 $\mu$L minute$^{-1}$, to maximize signal and reduce noise \citep{wolters2020}. Initial measurements were taken in positive ion mode. The instrument polarization was then switched to negative ion mode and allowed to re-equilibrate for 90 minutes. With the instrument in standby, the capillaries were flushed with 10 $\mu$L of CH$_3$OH and then the direct injection polyetheretherketone (PEEK) capillary replaced. The instrument was then mass calibrated again and another blank solution was taken in negative polarity before the Triton sample was measured in negative ion mode with the same instrument settings as above. 

\subsection{Data Analysis of Orbitrap MS with \texttt{idmol}}
After data acquisition and preliminary inspection with Thermo Fisher Scientific XCalibur software, we use custom IDL/Fortran software called \textit{idmol} \citep{horstthesis} to extract and analyze the mass spectral data. First, we compare the mass spectra of blank solution to that of the Triton analogue and remove contaminant peaks with intensities greater than 2$\times$10$^5$ in the Triton spectra. More recent work suggests that this method of blank subtraction can remove useful signal due to the behavior of the automatic gain control (AGC) of the Orbitrap (see Wolters et al., forthcoming), but in this case we do not remove many peaks and therefore expect minimal impact on our results (see Appendix, Figure \ref{fig:blank}). Next, we use \textit{idmol} to assign molecular peaks, where the program calculates all potential molecular combinations and then uses a series of decision trees based on user inputs regarding minimum N/C ratio, maximum number of oxygens, mass tolerance, and the nitrogen rule to make assignments of low-mass mass peaks. Above m/z 300, where non-unique mass peaks are no longer within the instrument's resolving power \citep{gautier2014orbitrap}, the program uses lower mass peak assignments to choose best fit assignments by eliminating potential molecules which substantially deviate from the average number of carbons, nitrogens, and double bond equivalent for the previous 20 assignments \citep{horstthesis}. 

As in previous analyses \citep{moran2020,vuitton2021}, we calculate the elemental composition of the tholin sample from Orbitrap measurements using \textit{idmol} peak assignments weighted by the spectral intensity. We perform a correction to the intensity weighted elemental composition by multiplying the lowest 10\% intensities by 10 to account for differing ionization yields of oxygen molecules, as was shown by \citet{horstthesis} to mitigate differences between Orbitrap and combustion derived elemental compositions. We also note that additional factors can introduce uncertainty into elemental ratios derived from Orbitrap results, such as the solubility of sample or the efficiency of ion transmission, in addition to the inherent error accounted for with our formula assignments. We also calculate the non-intensity weighted elemental composition, as ionization efficiency may prevent a direct proportionality to molecular concentration and impact the derived elemental ratios. For further discussion regarding ionization efficiencies of various chemical species, including as a function of positive and negative polarities, see \citet{vuitton2021}. 

\section{Results} \label{sec:results}
In this section, we present the results of the production rate, our elemental, mass spectrometry, and Fourier Transform Infrared (FTIR) spectroscopy analysis. Broadly, CO from the initial gas mixture profoundly influences the resulting composition and slightly alters the spectra of the solid particles. First, we discuss production rate of haze as a function of CO and CH$_4$ mixing ratios. Then we focus on the bulk composition as determined by both combustion and high resolution mass spectrometry (HRMS). Next we delve deeper into the composition and specific molecular identifications enabled by HRMS, and finally we examine the Triton tholin FTIR spectra for signs of the novel chemistry that may be observable by remote sensing.

\subsection{Production Rate of Triton Tholin Particles}
\label{sec:productionrates}
  
\begin{figure*}[hbpt!]
\centering
\includegraphics[angle=0,width=0.95\linewidth]{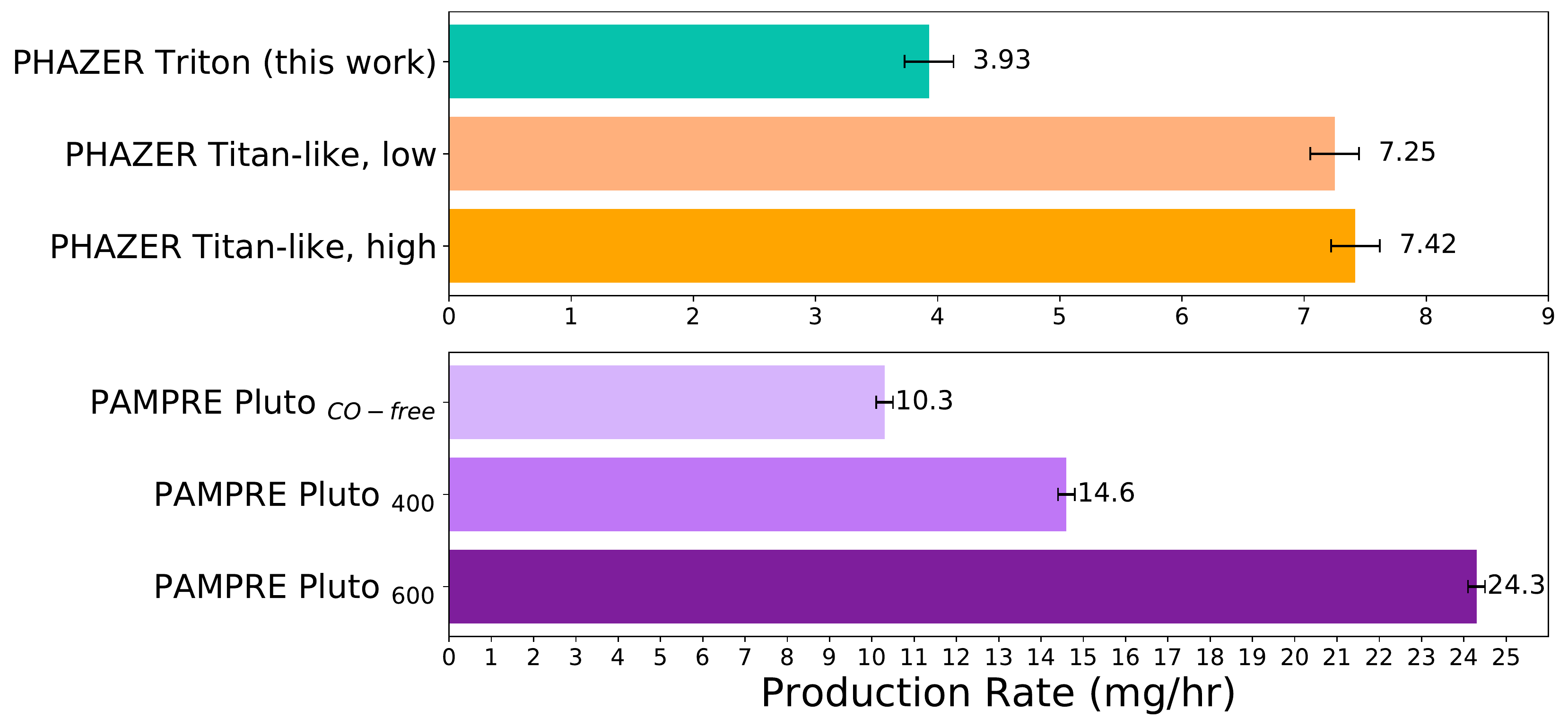}
\caption{Top: PHAZER production rates (in mg/hr) of the amount of solid produced from the Triton gas mixture (CO = 0.5\%; CH$_4$ = 0.2\%; top bar in turquoise) compared to our Titan-like tholin (CH$_4$ = 5\%; CO in varying mixing ratios from 0\% to 5\%; \citealt{he2017co}, middle and lower orange bars). Bottom: PAMPRE production rates from the room temperature Pluto-like tholin (P$_{CO-free}$: CO = 0\%, CH$_4$ = 1\%, top pale purple bar; P$_{400}$: CO = 500ppm; CH$_4$ = 1\%, middle purple bar; and P$_{600}$: CO = 500ppm, CH$_4$ = 5\%, lower dark purple bar) of \citet{JovanovicPlutoOrbitrap}. Note that due to differing experimental set-ups, the absolute production rates between the PHAZER and PAMPRE apparatus cannot be directly compared and have different x-axis limits. While CO alters the haze chemistry, higher methane mixing ratios generate larger amounts of haze material in both experimental set-ups.}
\label{fig:prodrate}
\end{figure*}

In Figure \ref{fig:prodrate}, we show the production rates of our Triton tholin experiment compared to that of our standard Titan gas mixture (5\% CH$_4$ in N$_2$) with varying CO percentages \citep{he2017co}. While the production rates measured by \citet{he2017co} found that varying CO from 0\% to 5\% with 5\% CH$_4$ did not impact the production rate within the limits of experimental error (0.20 mg/hr), our Triton experiment made markedly less haze, at only 3.93$\pm$0.2 mg/hr compared to the Titan-like production rates of 7.25-7.42 mg/hr. We also show the production rates (L. Jovanovi{\'c}, personal communication) of tholin experiments for Pluto gas composition from the PAMPRE experimental set-up \citep{szopa2006,alcouffe2009capacitively} at LATMOS, which used 500 ppm CO in N$_2$ with either 1\% CH$_4$ to simulate Pluto's atmosphere at altitude of 400 km or 5\% CH$_4$ for an altitude of 600 km, as well as a CO-free experiment with 99\% N$_2$ and 1\% CH$_4$ \citep{JovanovicPlutoOrbitrap}, all at room temperature. A description of the experimental conditions for both sets of experiments can also be found in Section \ref{sec:comparison} in Table \ref{table:exp_cond}. Like our Triton and Titan results, the Pluto-like results show a much higher production rate when the methane mixing ratio is higher (14.6 mg/hr with 1\% CH$_4$ compared to 24.3 mg/hr with 5\% CH$_4$.) While the differing experimental set-ups prevent a direct comparison between the absolute production rates seen in PHAZER and PAMPRE, a trend in the relative yields (i.e., normalizing the PAMPRE values to the PHAZER values as in Figure \ref{fig:prodrate}) is suggested. The much lower starting mixing ratio of methane in the Triton mixture (0.2\%) and low-altitude Pluto composition mixture (1\%) results in much less haze compared to that of our Titan experiments or high-altitude Pluto-like experiments with 5\% methane. Even more striking is the effect of removing CO from the system enitrely, as seen in the CO-free, 1\% CH$_4$ in N$_2$ experiment of \citet{JovanovicPlutoOrbitrap}. Despite the same amount of methane as in their Pluto$_{400}$ condition, their CO-free condition made just 10.3 mg/hr of solid. In fact, the relative yield of their CO-free experiment is lower than our Triton experiment which has an order of magnitude less CH$_4$, but an order of magnitude more CO.

Other investigations of N$_2$-CH$_4$ mixtures have shown that aerosol mass loading is always higher with 2\% CH$_4$ than with 0.1\% CH$_4$ using a spark discharge energy source, though the opposite is observed when the energy source is a UV lamp \citep{horstandtolbert2013}. Given the Triton and Titan PHAZER experiments, our glow discharge is thus likely more analogous to the spark discharge, where the larger methane gas mixing ratio allows for the production of more solid material. In a later experiment with the same set-up, the total abundance of gas phase products also increased with higher methane mixing ratios in the initial gas mixture \citep{horst2018titan}. PHAZER exoplanet atmospheric studies have also shown that higher initial methane mixing ratios can increase solid material production \citep{horst2018production}, though the third highest observed production rate of those experimental conditions did not include an initial gas phase inventory of methane at all. Additional Pluto-like composition experiments (of 500 ppm CO in N$_2$) observed significantly thicker thin films were produced under 5\% CH$_4$ conditions than with 0.5\% or 1\% CH$_4$ \citep{jovanovic2021opticalconstants}. We also note that all these experiments have been performed at highly varying temperature conditions (90 K to 800 K), adding a further layer of complication in extrapolating trends. Prior work has shown that Titan's methane mixing ratio appears to most efficiently convert gas precursors to tholin solids due to competing solid growth mechanisms when CH$_4$ increases beyond this percentage  \citep{trainer2006,sciammaobrien2010}. The exact percentage of methane represented in the lab depends on the optical depth of the set-up in question \citep{trainer2006}.

While we clearly demonstrate in later sections that the haze chemistry changes under the influence of very small amounts of CO, the literature suggests CH$_4$ may play a larger part in the ultimate aerosol mass loading. Current trends in the literature when combining CO and CH$_4$ remain tentative or contradictory and could benefit from further study \citep[see e.g.,][]{horstandtolbert2014}. Early Earth studies have also examined the effect of combining methane with carbon dioxide, finding that a certain amount of CO$_2$ with CH$_4$ actually increases particle production up to a CH$_4$/CO$_2$ ratio of 1 before decreasing \citep{trainer2006}, though particles are still produced even when the CO$_2$ ratio is much higher \citep{dewitt2009}. Therefore, while we identify the methane mixing ratio as the major driver behind increased production, it remains unclear if any carbon source in high enough quantities could serve to generate substantial haze (see, e.g, \citealt{horst2018production,He2018,he2020}). This is underscored with the marked decrease in production rate when CO is removed entirely with a low methane mixing ratio as in \citet{JovanovicPlutoOrbitrap}'s CO-free Pluto experiment. Clearly the balance of the two carbon sources is critical to the ultimate haze mass loading.

\subsection{Composition of Triton Tholin Particles} \label{sec:compositionresults}

\subsubsection{Bulk Composition and Differences in Ion Polarities}
We observe first that despite CO being present at only a 0.5\% mixing ratio in the initial gas mixture, the bulk solid is approximately 10\% oxygen. From the combustion analysis, we determine that H, O, C, and N are present at 5.29$\pm$0.06\%, 10.3$\pm$0.3\%, 43.7$\pm$0.4\%, and 40.7$\pm$0.4\% respectively. The bulk composition as calculated from the combined positive and negative ions of the soluble fraction of the tholin from Orbitrap is 5.4$\pm$0.7\% H, 9.8$\pm$4.3\% O, 47.4$\pm$1.7\% C, and 37.4$\pm$2.8\% N. The bulk composition from the Orbitrap agrees to within 2\% or better with the combustion analysis. This suggests that the soluble fraction of the Triton tholin, as measured from Orbitrap HRMS, is fairly representative of the global sample from an elemental perspective. Previous work has shown that the soluble and insoluble fractions of tholin can be quite different chemically \citep{maillard2018solubility}. The similarity of our elemental analyses as calculated from Orbitrap and combustion demonstrates that this effect is likely minor for our Triton tholin, though it is possible that even with the same elemental composition, the chemical composition and structure can still be quite different between the soluble and insoluble fractions.

\begin{figure*}[hbt!]
\centering
\includegraphics[angle=0,width=0.95\linewidth]{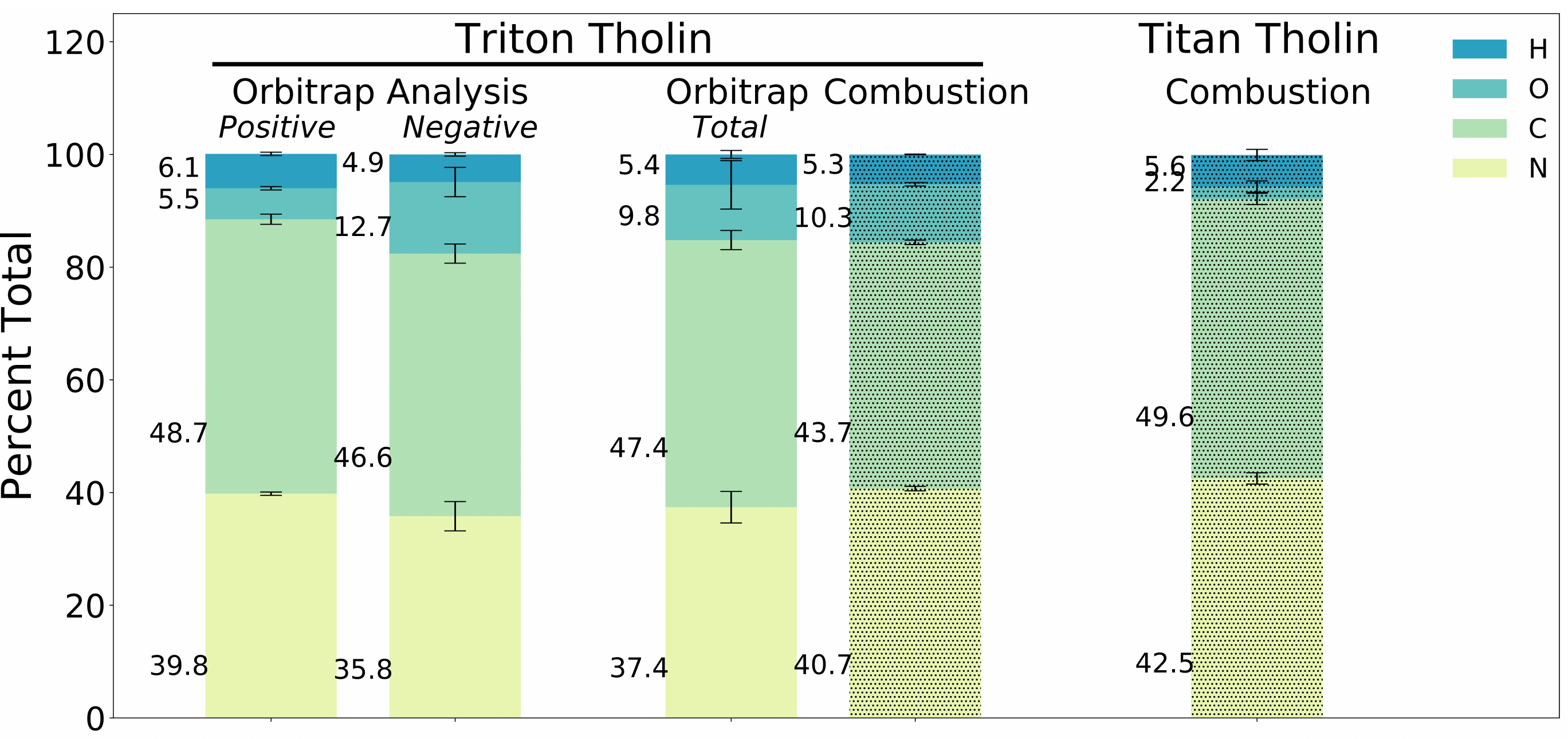}
\caption{Average elemental composition of Triton tholin from positive, negative, and combined ions, as determined by Orbitrap MS analysis (left three bars), combustion analysis (fourth bar), and comparison to standard PHAZER Titan values from combustion analysis (right). Triton tholin more strongly incorporates oxygen, apparently through carbon depletion. Hatching on the right two bars indicates these results come from combustion analysis.}
\label{fig:elemental}
\end{figure*}

As seen in previous work \citep{horstthesis}, oxygen is primarily detected in the negatively ionized products from the Orbitrap, which is consistent with the identification of carboxylic acid groups (discussed further in the following subsections) which are efficient proton donors \citep{vuitton2021}. The bulk values obtained for negative, positive, and combined ions from Orbitrap, as well as from combustion analysis, are shown in Figure \ref{fig:elemental}.

Titan tholin produced without CO from the same experimental set-up contains $\sim$2\% oxygen (likely due to minor water adsorption during measurements) compared to the Triton tholin with $\sim$10\% oxygen (Figure \ref{fig:elemental}). Previous PHAZER CO experiments \citep{he2017co} have included up to 5\% CO in N$_2$-CH$_4$ mixtures, yet never observe more than 8\% oxygen in the solid and indeed measure only 5\% oxygen in mixtures with starting CO at 0.5\% as in this work. Importantly, these prior experiments were conducted with constant CH$_4$ gas mixing ratios of 5\% and did not include a case where CO $>$ CH$_4$. Thus our Triton haze analogues demonstrate how much more strongly oxygen can be incorporated into the solid when CO $>$ CH$_4$ in the starting gas mixture. Similar results from CO$_2$-CH$_4$ experiments also show a marked chemical transition when CH$_4$/CO$_2$ reaches unity \citep{trainer2006}. Hydrogen contents remain at approximately 5\% across varying CO/CH$_4$ starting ratios \citep{he2017co}, and our Triton results here support this trend, which is also seen when CO$_2$ and H$_2$ ratios are varied with methane \citep{dewitt2009}. Our elemental analysis results in a bulk N/C ratio for the Triton tholin of 0.93, which is higher than that of ``standard'' PHAZER Titan tholin at N/C = 0.85. This result is surprising, considering that past work with N$_2$-CH$_4$-CO mixtures has postulated that the competition between N$_2$ and CO chemistry results in oxygen incorporation primarily at the expense of nitrogen incorporation \citep{he2017co,JovanovicPlutoOrbitrap}. Here, instead, carbon appears depleted with respect to the Titan tholin structure. While this could simply result from CO $>$ CH$_4$ in the gas mixture, this result is more in line with previous work that explored the addition of CO$_2$ and O$_2$ in N$_2$-CH$_4$ atmospheres \citep{horst2018earlyearth}, which found increasingly oxidized gas mixtures increased nitrogen fixation in the solid.
A previous Triton experiment also showed stronger nitrogen incorporation compared to Titan tholin \citep{mcdonald1994titanandtritontholin}, though that experiment contained no CO and simply reflects the lower methane mixing ratio of the starting gas mixture. For PHAZER Titan tholin, the C/O ratio of the bulk solid is 22.5 but the Triton tholin's bulk C/O is only 4.2. Even the previous PHAZER CO experimental series with the highest starting CO/CH$_4$ gas ratio (1:1) resulted in a C/O ratio for the tholin of $\sim$6, considerably higher than we observe for the Triton tholin. In our Triton case, it remains unclear therefore whether the higher nitrogen and oxygen contents we see derive from the lower total carbon abundance in the starting gas mixture or the presence of CO in particular.

\subsubsection{Molecules in the Triton Sample from HRMS}

\begin{figure*}[hbt!]
\centering
\includegraphics[angle=0,width=0.95\linewidth]{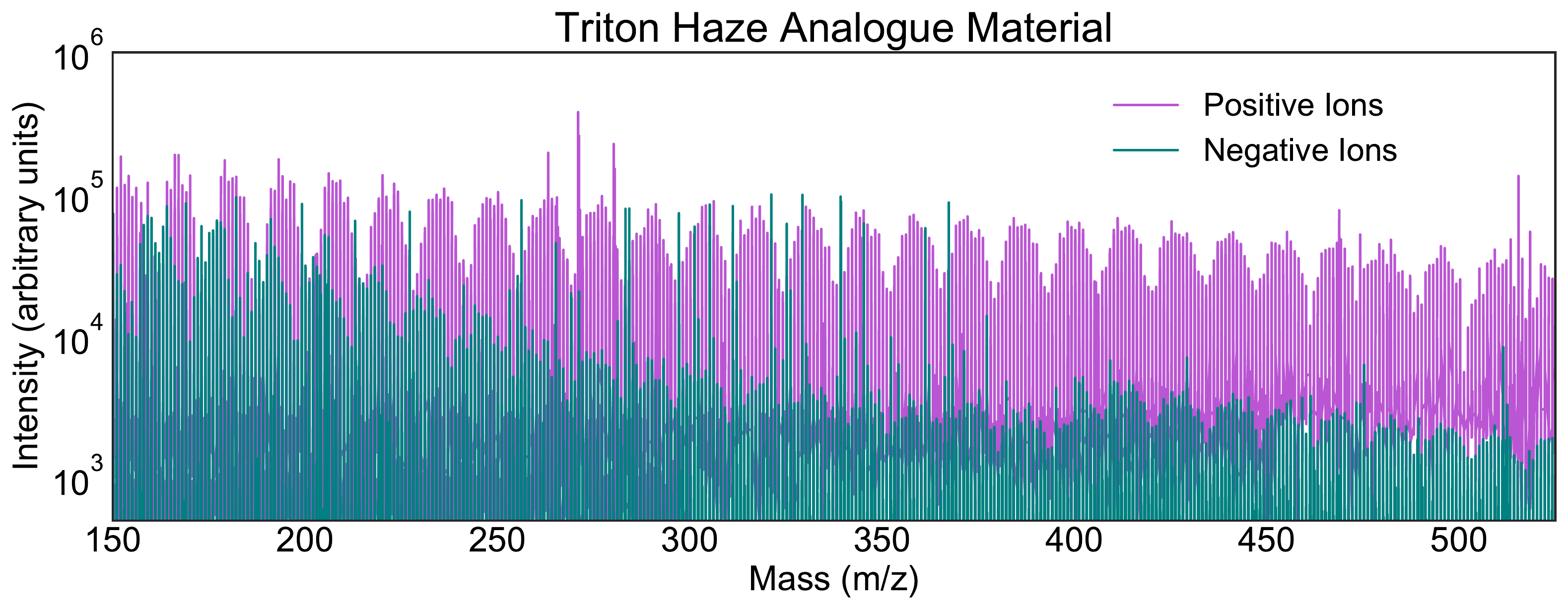}
\caption{Positive (magenta) and negative (teal) ion mode mass spectra of the Triton haze analogue particles from m/z 150 to 525. Due to differing ionization efficiencies, the negative ion mode intensities are systematically lower. Peak groupings of $\sim$13.5 amu repeat across the observed mass range.}
\label{fig:mass_spec}
\end{figure*}

Following our characterization of the bulk Triton haze analogue sample and its bulk soluble fraction, we examined the mass spectral data for specific molecular formulas of interest. The positive and negative mass spectra are shown in Figure \ref{fig:mass_spec}. We detect thousands of individual peaks across both ionization polarities, indicating the high complexity of the Triton tholin as is the case with its Titan, Pluto, and exoplanet counterparts \cite[e.g.,][]{gautier2016hplcorbitrap,he2017co,JovanovicPlutoOrbitrap,moran2020}. Distinct periodicity is observed in the peak groupings, with a repeating pattern between 13 and 14 amu across both the positive and negative ions. Positive ions have peak group periods that average to 13.57 amu; negative ions cluster in peak groupings with average spacing of 13.66 amu. These values are quite close to the 13.5 amu observed by \citet{horstthesis} and consistent with the frequently observed 12 to 16 amu spacing (averaging to 14 amu) in Titan and Pluto tholin produced in various set-ups \citep{imanaka2004,cabletholinreview2012,gautier2014orbitrap,gautier2017,JovanovicPlutoOrbitrap}. These peak spacings display quite regular but not exact amu periodicity. Therefore, these groupings likely indicate repeating chemical units of co-added and substituted monomers (such as HCN, CH$_2$, etc.) \citep{gautier2017}.

\begin{table}[hbt!]
\caption{Molecular Formulas Identified by Orbitrap High Resolution Mass Spectrometry}
\begin{tabular}{@{}llllllll@{}}
\toprule
\multicolumn{1}{l}{m/z $\pm$} & \multicolumn{1}{l}{$\Delta$ppm} & \multicolumn{1}{l}{Ion Formula} && \multicolumn{1}{l}{Intensity} && 
\multicolumn{1}{l}{Potential Molecule} &
\multicolumn{1}{l}{In PHAZER Titan Sample?} \\ \midrule
90.0312 & 0.18 & C$_3$H$_6$O$_3^-$ && 9.4 x 10$^3$ && Glyceraldehyde & \textbf{No} \\
111.043 & -0.45 & C$_4$H$_5$N$_3$O$^+$ && 1.8 x 10$^4$ && Cytosine & Yes \\
112.027 & 0.18 & C$_4$H$_4$N$_2$O$_2^-$ && 6.9 x 10$^2$ && Uracil & Yes \\
126.043 & 0.21 & C$_5$H$_6$N$_2$O$_2^-$ && 8.0 x 10$^2$ && Thymine & Yes \\
135.054 & -0.97 & C$_5$H$_5$N$_5^+$ && 1.6 x 10$^4$ && Adenine & No \\
 & 0.43 & C$_5$H$_5$N$_5^-$ && 9.3 x 10$^3$ &&& Yes \\
151.049 & -1.56 & C$_5$H$_5$N$_5$O$^+$ && 9.0 x 10$^3$ && Guanine & No \\
& -0.10 & C$_5$H$_5$N$_5$O$^-$ && 7.4 x 10$^4$ &&& Yes\\
151.063 & 0.90 & C$_8$H$_9$NO$_2^-$ && 2.5 x 10$^3$ && 2-phenylglycine & No \\
152.033 & -2.27 & C$_5$H$_4$N$_4$O$_2^-$ && 1.3 x 10$^3$ && Xanthine & Yes \\
155.069 & -1.45 & C$_6$H$_9$N$_3$O$_2^+$ && 9.2 x 10$^3$ && Histidine & Yes \\
 & -0.41 & C$_6$H$_9$N$_3$O$_2^-$ && 1.0 x 10$^4$ &&& Yes \\
156.065 & -0.74 & C$_5$H$_8$N$_4$O$_2^+$ && 1.2 x 10$^4$ && 1,2,4-triazole-3-alanine & Yes \\
& -0.04 & C$_5$H$_8$N$_4$O$_2^-$ && 1.4 x 10$^4$ &&& Yes \\
160.048 & 0.57 & C$_5$H$_8$N$_2$O$_4^-$ && 4.1 x 10$^3$ && Thymine glycol & No \\
167.069 & -0.68 & C$_7$H$_9$N$_3$O$_2^-$ && 6.4 x 10$^3$ && $\beta$-pyrazinyl-L-alanine & Yes \\
169.085 & 0.18 & C$_7$H$_{11}$N$_3$O$_2^+$ && 1.4 x 10$^4$ && 3-methylhistidine & Yes \\
 & -0.78 & C$_7$H$_{11}$N$_3$O$_2^-$ && 6.0 x 10$^3$ &&& Yes \\
171.064 & 0.15 & C$_6$H$_{9}$N$_3$O$_3^-$ && 3.2 x 10$^3$ && $\beta$-hydroxyhistidine & Yes \\
172.096 & 0.05 & C$_6$H$_{12}$N$_4$O$_2^+$ && 2.7 x 10$^4$ && Iminoarginine & Yes \\
 & 0.55 & C$_6$H$_{12}$N$_4$O$_2^-$ && 1.2 x 10$^3$ &&& No \\
182.080 & 0.72 & C$_7$H$_{10}$N$_4$O$_2^+$ && 1.8 x 10$^4$ && Lathyrine & Yes \\
 & 0.54 & C$_7$H$_{10}$N$_4$O$_2^-$ && 1.4 x 10$^4$ &&& Yes \\
193.074 & 0.32 & C$_{10}$H$_{11}$NO$^-$ && 1.3 x 10$^3$ && Phenylglycine, m-acetyl & No \\
206.080 & 0.76 & C$_{9}$H$_{10}$N$_4$O$_2^-$ && 6.2 x 10$^3$ && Benzotriazolylalanine & Yes \\
226.107 & -0.26 & C$_{9}$H$_{14}$N$_4$O$_3^-$ && 2.8 x 10$^3$ && Alanylhistidine & Yes \\
246.100 & -1.85 & C$_{13}$H$_{14}$N$_2$O$_3^-$ && 1.6 x 10$^3$ && Acetyltryptophan & No \\
\bottomrule
\end{tabular}
\\ \textit{Note.} the final column notes whether this formula is seen in Titan sample produced with the PHAZER set-up under CO-free conditions.
\label{table:molecular_formulas}
\end{table}
We report in Table \ref{table:molecular_formulas} a small subset of the thousands of molecular formulas of species identified from Orbitrap MS with intensities $\geq$10$^3$. These molecular formulas are consistent with those of various prebiotic molecules, though we stress that without structural information (which is not possible with the measurements conducted in this work), we cannot confirm that the formula we identify is in fact the isomer of the prebiotic species indicated in Table \ref{table:molecular_formulas}. We detect the formulas for all biologically relevant nucleotide bases (adenine, C$_5$H$_5$N$_5$; cytosine, C$_4$H$_5$N$_3$O; guanine, C$_5$H$_5$N$_5$O; thymine, C$_5$H$_6$N$_2$O$_2$; and uracil, C$_4$H$_4$N$_2$O$_2$) and one non-biologic nucleotide base (xanthine, C$_5$H$_4$N$_4$O$_2$). We note that while we include the formulas for uracil and thymine because they represent the remaining nucleotide bases, they are present only at 10$^2$ intensities. Each of these species has been previously identified and confirmed through alternate measurement techniques in prior tholin produced from N$_2$-CH$_4$-CO mixtures \citep{horst2012formation,sebree2018}, raising the likelihood that the formulas we identify in the Triton tholin could be the prebiotic species in question. We also identify numerous formulas consistent with isomers of amino acids and their derivatives. Nearly all of these formulas are also present as at least one ion in HRMS data of tholin produced under CO-free Titan conditions with the PHAZER set-up, as indicated in the far right column of Table \ref{table:molecular_formulas}. Because we identify most of these oxygenated molecules in the CO-free Titan sample as well as in our Triton sample, we cannot assume they are formed during sample synthesis, but must consider that they could be the result of contamination. Despite this possibility, many of these formulas have been unambiguously identified from CO-containing Titan experiments \citep[e.g.,][]{horst2012formation} by using $^{18}$O labeled CO, which we could employ in future studies to eliminate contamination concerns.

In contrast to these ambiguous detections, and most notably, we detect here the formula C$_3$H$_6$O$_3$ with a potential isomer of glyceraldehyde, the simplest possible monosaccharide. This species' formula is unique to the Triton tholin among all N$_2$-CH$_4$-CO experiments to which we compare our results. Only prior tholin work considering significantly more oxidizing atmospheres has previously observed sugar formulas \citep{moran2020}. Again, this result underscores the impact of a starting CO/CH$_4$ mixing ratio over unity. If confirmed, the existing co-presence of nucleotide bases, amino groups, and sugar molecules in  the complex solids that make up this haze should significantly advance Triton as a target of astrobiological interest. For each identified formula in Table \ref{table:molecular_formulas}, we confirm that it is not present in the Orbitrap data of the blank solvent above the noise level (intensities $\geq$ 10$^1$). 

\begin{figure*}[hbpt!]
\centering
\includegraphics[angle=0,width=0.95\linewidth]{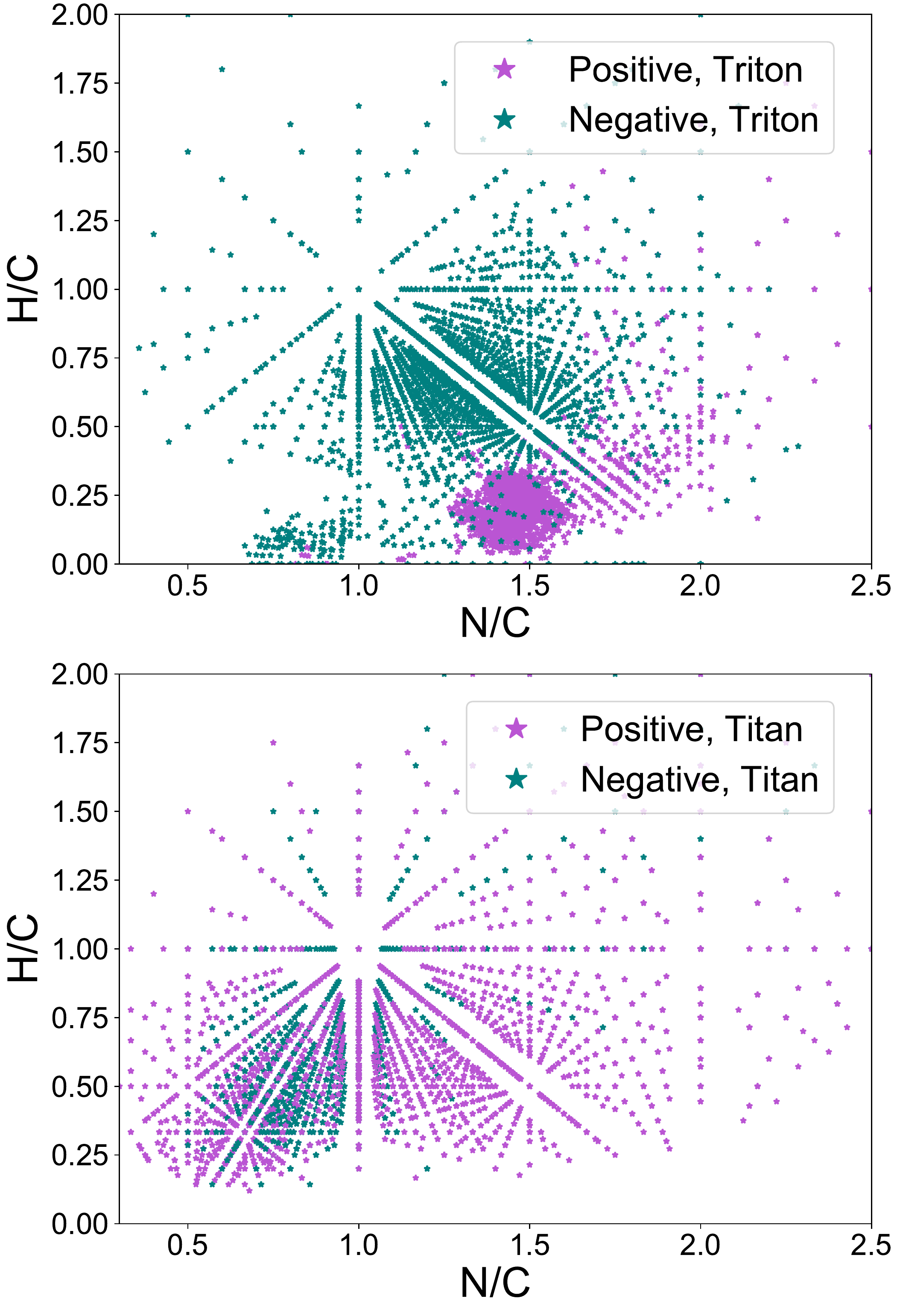}
\caption{van Krevelen diagrams of the PHAZER Triton haze analogues (top) and PHAZER Titan haze analogues (bottom) in terms of H/C vs N/C, as is common in other tholin studies. The molecular character clearly differs between negative and positive ionizations, underscoring the need for both to understand the sample's chemistry. The nitrogen incorporation also clearly differs between the Titan sample and the Triton sample produced from a gas mix with CO dominant over CH$_4$.}
\label{fig:vk}
\end{figure*}

\begin{figure*}[hbpt!]
\centering
\includegraphics[angle=0,width=0.95\linewidth]{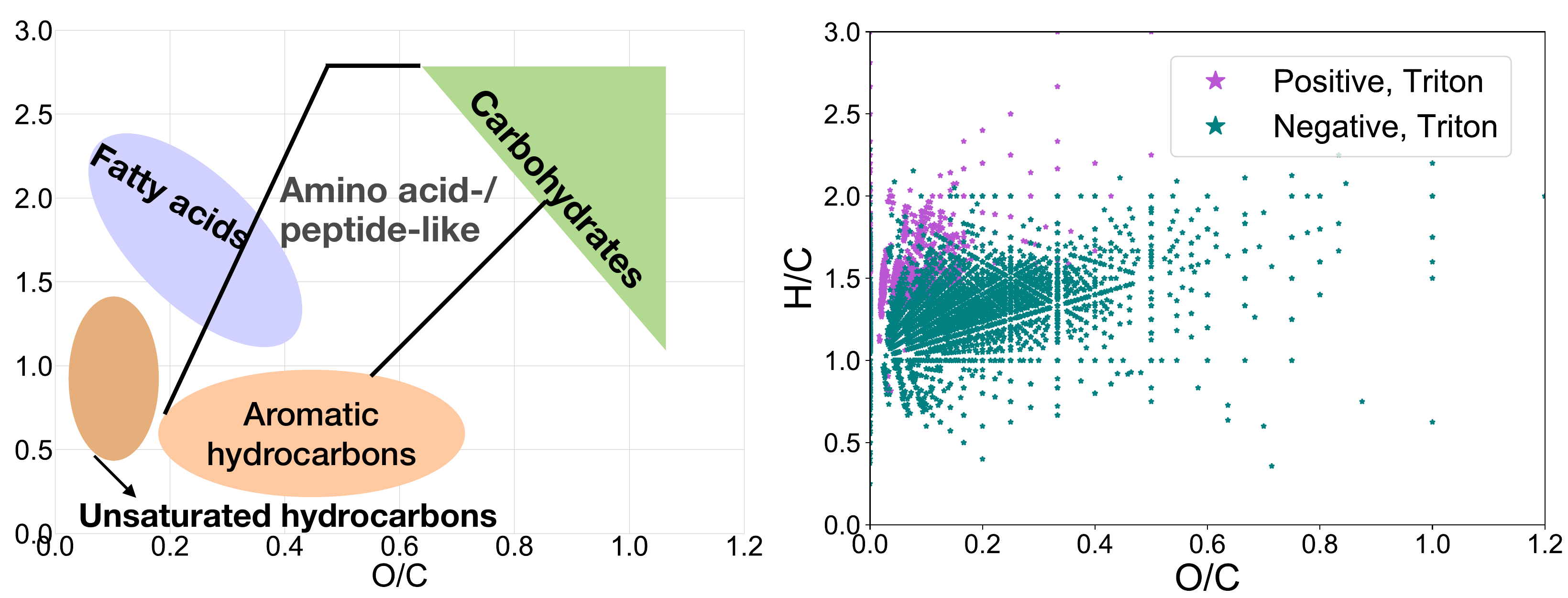}
\caption{LEFT: An empty van Krevelen diagram showing H/C vs O/C with shaded, labeled regions denoting where particular molecular functional groups tend to cluster, following \citet{Ruf2018VKastro}. RIGHT: van Krevelen in terms of H/C vs O/C for our Triton sample. The Triton sample clusters strongly in the regions for fatty acids and amino acid-like molecular groups.}
\label{fig:vk_labels}
\end{figure*}

In Figure \ref{fig:vk}, we present van Krevelen diagrams \citep{Krevelen1950GraphicalstatisticalMF} using H/C vs. N/C ratios of the identified molecules, which are a modification frequently used in the planetary literature to study complex organic materials like tholin \cite[e.g.,][]{horstthesis,gautier2014orbitrap,JovanovicPlutoOrbitrap}. In Figure \ref{fig:vk_labels}, we show H/C vs. O/C ratios of the identified molecules, which are widely used to understand the structures of complex organic matter on Earth \cite[e.g.,][]{Kim2003VK}. We separate the molecular ratios by whether they were seen in positive or negative ion mode. The difference in clustering between ion polarities clearly demonstrates that both are needed to understand the full chemical nature of the tholin's soluble fraction, which is frequently only a subset of the global molecular complexity \citep{maillard2018solubility}. Strongly oxygenated molecules clearly favor negative ionization while more heavily nitrogenated molecules increase in positive ionization mode. In H/C vs N/C space, we compare our results to PHAZER Titan tholin produced without CO. The Titan van Krevelen diagrams use molecular assignments performed with slightly different software, \textit{Attributor}, \citep{orthousdaunay2020}, but which has been benchmarked against \textit{idmol} and reliably produces comparable results at masses m/z $\leq$300 \citep{horstthesis,bonnet2013compositional}.

The Triton tholin reach high O/C ratios across all H/C ratios. Oxygen species are always more prevalent in the negative ionization data and consequently reach larger O/C ratios. We also show in Figure \ref{fig:vk_labels} an empty van Krevelen diagram that shows the regions in which particular molecular functional groups fall, for reference. The Triton tholin distinctly cluster above an H/C of 1.0, the demarcation of oxidation reactions and more highly unsaturated species. From the elemental analysis, however, we know that the overall hydrogen content is comparable between the the Triton and Titan samples, reiterating our previous interpretation that oxygen is incorporated into the solid at the expense of carbon. The Triton tholin exhibit a shift away from simple unsaturated hydrocarbons to long chain carboxylic acids (fatty acids), bounded by H/C ratios of 1.0 -- 2.0 and an O/C ratio $\leq$ 0.4. Additionally, the Triton tholin displays a large clustering of molecules consistent with amino acid and peptide-like species (1.0 $\geq$ H/C $\leq$ 2.0; 0.2 $\geq$ O/C $\leq$ 0.8). Finally, the Triton tholin, due to its increased oxidation, shows greater clustering above H/C of 1.5 and O/C of 0.8, where carbohydrate species fall on van Krevelen diagrams \citep{Ruf2018VKastro}.

In H/C vs. N/C space, the Titan tholin is relatively similar between positive and negative ionization, but the Triton tholin is dramatically different. While the negatively ionized molecules from the Triton tholin are qualitatively similar to Titan in this visualization, the nitrogenation of the Triton tholin is clearly enhanced above that of the Titan tholin as seen in the intense clustering of molecules above an N/C of 1.0. These results are also in strong contrast to the results of \citet{JovanovicPlutoOrbitrap}, who studied varying CO-containing N$_2$-CH$_4$ mixtures for Pluto atmospheric composition. \citet{JovanovicPlutoOrbitrap} found, like with our Titan tholin, a preference for molecules with N/C ratios less than one, suggesting that oxygen is incorporated into the solid at the expense of nitrogen. As discussed in our elemental analysis results, we find instead that carbon depletion is associated with the addition of oxygen atoms into the tholin while nitrogen comparatively increases. These results strongly suggest an alternative chemical pathway is at work in the production of the solids when CO gas $\geq$ CH$_4$. The shift to a more oxidizing atmosphere increasing nitrogen fixation has precedence in early Earth studies with CO$_2$ and O$_2$ \citep{horst2018earlyearth,gavilan2018organicaerosols}. Morever, unlike our PHAZER Titan tholin, the positively ionized molecules in our PHAZER Triton tholin are remarkably tightly confined between N/C ratios of 1.2 and 1.5 and H/C ratios of 0 to 0.5. Combined with the increase in unsaturation and oxidation, this tight clustering is likely demonstrating that hydrogen atoms preferentially bond to nitrogen and oxygen atoms. As postulated in \citet{horstandtolbert2014} and \citet{he2017co}, these reactions may occur due to oxygen radicals from CO efficiently removing hydrogen atoms and molecules from CH$_4$ along the pathway to solid haze formation.

\subsection{Transmission and Reflectance Spectra of the Triton Tholin}

\subsubsection{Functional Groups from VIS to NIR}

\begin{figure*}[hbt!]
\centering
\includegraphics[angle=0,width=0.95\linewidth]{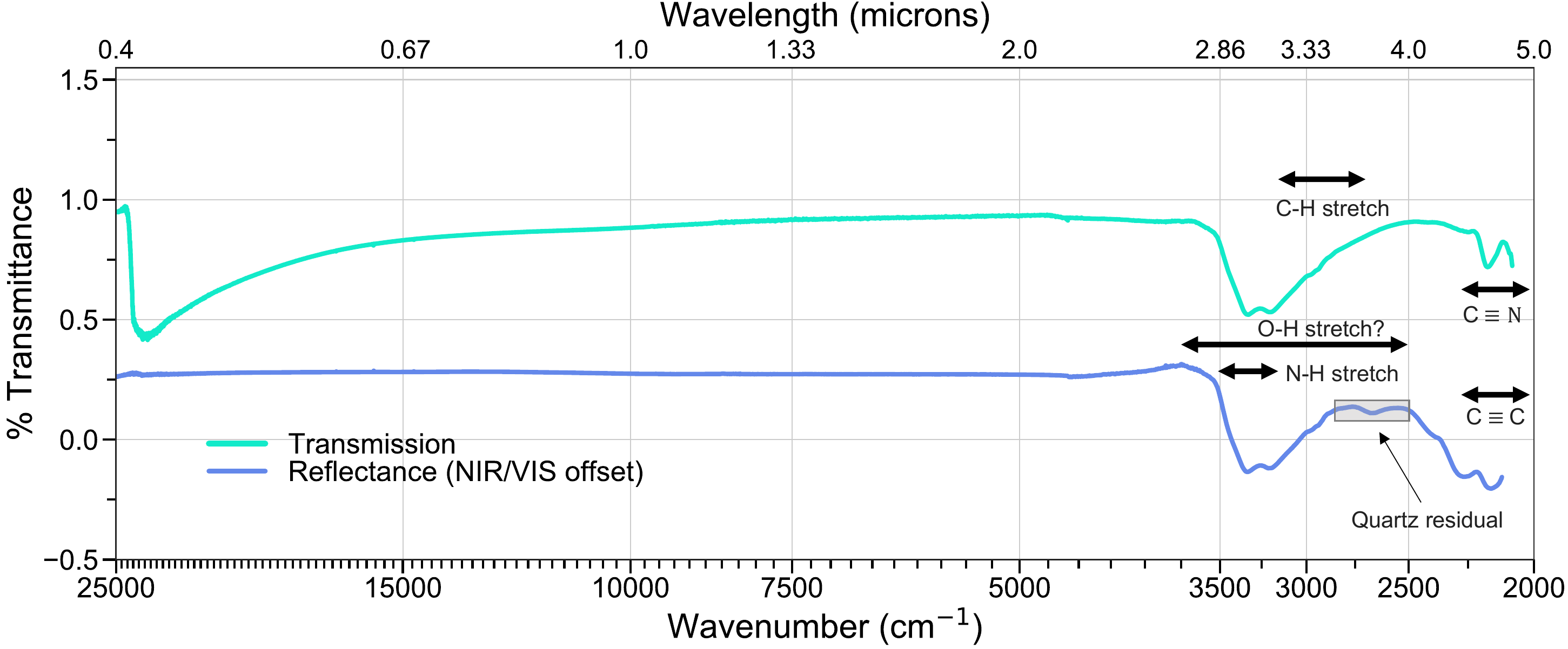}
\caption{Transmittance (teal) and reflectance (periwinkle) of the Triton tholin. A downward slope is visible from 15000 cm$^{-1}$ in transmission and features from O-H, N-H, C-H, C=C, and C$\equiv$N are present from 4000 to 2000 cm$^{-1}$ in both transmittance and reflectance spectra. The reflectance spectrum has been offset from the transmittance spectrum in the NIR by 0.4 for clarity. Note that the reflectance spectrum contains a residual absorption feature at $\sim$2600 cm$^{-1}$ likely from the quartz substrate.}
\label{fig:transrefl}
\end{figure*}

We show in Figure \ref{fig:transrefl} the transmittance and reflectance spectra of the Triton tholin from visible to near-infrared wavelengths, 0.4 to 5 $\mu$m (25000 cm$^{-1}$ to 2000 cm$^{-1}$). In visible wavelengths, the most noticeable characteristic of the tholin is its sharp downward slope from 15000 cm$^{-1}$ in transmission, which is observable to the eye as the tholin's brown-ish orange color, well known from Titan tholin studies. In the near-IR, we attribute the variety of features observed to a combination of N-H, C-H, C$\equiv$N, C=C, and O-H bonds. Amine (N-H and N-H$_2$) features are well described in Titan and Pluto tholin in the 3500 cm$^{-1}$ to 3000 cm$^{-1}$ regions \cite[e.g.,][]{imanaka2004,JovanovicPlutoOrbitrap}, but given the prevalence of O species identified in the mass spectral data and combustion analysis, we assign a subset of these features tentatively to O-H alcohol bonds in addition to amines. There is slight deepening and broadening of the amine feature out to 2500 cm$^{-1}$, which could derive from carboxylic acid species with O-H bonds, which would be consistent with the clustering we describe above in the van Krevelen diagrams.  Around the 3000 cm$^{-1}$ region, we identify a feature consistent with various C-H bonds, namely alkenes and alkanes. Between 2300 and 2000 cm$^{-1}$, we see some combination of nitrile, alkyne, and imine functional groups. Overall, this plethora of features underscores the complex nature of tholin. We provide a complete list of potential attributions of the spectra in Table \ref{table:spectra}. We also observe some weak features likely indicative of adsorption from ambient atmosphere or residuals in the spectrometer ($\sim$1 hPa) when the sample is briefly exposed to air (water at 3900-3700 cm$^{-1}$ and carbon dioxide at 2370 cm$^{-1}$), which is also seen in Pluto-like tholin \citep{JovanovicPlutoOrbitrap}.

Unfortunately, due to our quartz substrate which strongly absorbs in the mid-IR, we are not sensitive to wavelengths beyond 5 $\mu$m (2000 cm$^{-1}$), where carbonyl groups (C=O) have prominent spectral features. While other substrates (KBr, Si, etc.) would enable better mid-IR spectroscopy, these materials still require vetting for chemical non-reactivity in our experimental set-up, which is the subject of future work. With quartz, the only O bonds we can readily detect are O-H bonds, which could be alcohols or carboxylic acids. Both of these have clusters in the relevant regions of our van Krevelen diagrams (Figure \ref{fig:vk_labels}), but considerably more molecules fall in the carboxylic acid region (where the line with intercept at H/C = 2.0 has a slope of -1) than in the alcohol region (where the line with intercept at H/C = 2.0 has a slope of 0). Overall, the spectral features we are able to observe at wavenumbers above 2000 cm$^{-1}$ are consistent with the functional groups we identify from HRMS.

\begin{table}[hbt!]
\caption{Spectral Features and Corresponding Functional Groups Identified in the Triton Sample.}
\begin{tabular}{@{}llll@{}}
\toprule
\multicolumn{1}{c}{\textbf{\begin{tabular}[l]{@{}l@{}}Frequency \\ (cm$^{-1}$)\end{tabular}}} & \multicolumn{1}{c}{\textbf{\begin{tabular}[l]{@{}l@{}}Wavelength\\ ($\mu$m)\end{tabular}}} & \multicolumn{1}{c}{\textbf{\begin{tabular}[c]{@{}c@{}}Potential functional group\\ (bond, type)\end{tabular}}} & \multicolumn{1}{c}{\textbf{\begin{tabular}[c]{@{}c@{}}Feature \\ (extent, strength)\end{tabular}}} \\ \midrule
3900 - 3700 & 2.56 - 2.7 & H$_2$O adsorption (contaminant) & broad, weak \\ \addlinespace
3560 - 2889 & 2.8 - 3.46 & N-H stretching (amine) & broad, strong \\ \addlinespace
3337 & 2.99 & \begin{tabular}[c]{@{}l@{}}--NH$_2$ asymmetric stretching,\\  --NH-- stretching (amines)\end{tabular} & sharp, strong \\ \addlinespace
3260 & 3.07 & N-H stretching (amine) & sharp, strong \\ \addlinespace
3200 - 3190 & 3.12 - 3.13 & \begin{tabular}[c]{@{}l@{}}--NH-- stretching,\\  overtone of --NH$_2$ bending (amines)\end{tabular} & sharp, strong \\ \addlinespace
3500 - 2500 & 2.86 - 4.0 & \begin{tabular}[c]{@{}l@{}}O-H stretching (alcohols)?,\\ O-H stretching (carboxylic acids)?\end{tabular} & broad, strong \\ \addlinespace
3004 - 2906 & 3.32 - 3.44 & \begin{tabular}[c]{@{}l@{}}--CH$_3$, \\ --CH$_2$-- asymmetric stretching \\ (alkenes, alkanes)\end{tabular} & shoulder, weak \\ \addlinespace
2370 & 4.21 & CO$_2$ residual (contaminant) & shoulder \\ \addlinespace
2340 - 2215 & 4.27 - 4.51 & -C$\equiv$N stretching (nitrile) & broad, weak \\ \addlinespace
2210 & 4.52 & \begin{tabular}[c]{@{}l@{}}C$\equiv$C stretching (alkyne),\\ -C$\equiv$N stretching (aromatic nitrile)\end{tabular} & sharp, strong \\
\addlinespace
2145 & 4.66 & \begin{tabular}[c]{@{}l@{}}--C$\equiv$N stretching (nitrile), \\ --N=C=N-- (carbodiimide),\\  CO fundamental\end{tabular} & shoulder \\ \addlinespace
\bottomrule
\end{tabular}
\label{table:spectra}
\end{table}
\subsubsection{Comparison to Other Tholin Spectra}
\label{sec:comparison}

We compare our measured transmission spectrum of Triton tholin to a collection of literature-reported values for tholin generated from nitrogen/methane atmospheres. We do not report every single tholin experiment to date, but select experiments which contain CO or have a Triton-focus. We also generally include only those data published originally as spectra, rather than including each set of published optical constants. We make an exception for the work of Khare et al. due to their widespread use and for \citet{jovanovic2021opticalconstants} as their Pluto-like experimental mixtures are otherwise the most similar to our Triton mixture. The Titan optical constants of \citet{khare1984}, the Triton-specific optical constants of \citet{khare1994triton}, and the Pluto-like optical constants of \citet{jovanovic2021opticalconstants} provide the complex refractive indices, \textit{n} and \textit{k}, which they derive from their measured spectra. To convert these values back to spectra for comparison here, we follow the method from \citet{khare1984} given by

\begin{equation}
    R = \frac{(n-1)^2 + k^2}{(n+1)^2 + k^2}
\end{equation}

\begin{equation}
    T = (1 - R)e^{\lambda/4t\pi}e^{-k}
\end{equation}    
where \textit{R} and \textit{T} are the reflectance and transmittance, respectively, \textit{n} is the real refractive index, \textit{k} is the imaginary refractive index, $\lambda$ is the wavelength in microns, and \textit{t} is the film thickness, which in the region of interest (0.4 $\mu$m - 5 $\mu$m) is reported in each publication. We affirm that the reflectance spectrum we derive from the given complex refractive indices in \citet{khare1984} is identical to their Fig. 5. The full set of \textit{n} and \textit{k} from \citet{khare1984} and \citet{jovanovic2021opticalconstants} are available in the literature, but for completeness we include these values from \citet{khare1994triton} in Table \ref{table:khare94} in \ref{sec:appendix}.

A general summary of the experimental conditions used to generate each set of tholin spectra discussed here is found in Table \ref{table:exp_cond}. Note that a more detailed discussion of most experimental conditions, as well as the wider range of experiments used to measure optical constants of (Titan) tholin, can be found in the review by \citet{brasse2015}.

\begin{figure*}[hbtp!]
\centering
\includegraphics[angle=0,width=0.99\linewidth]{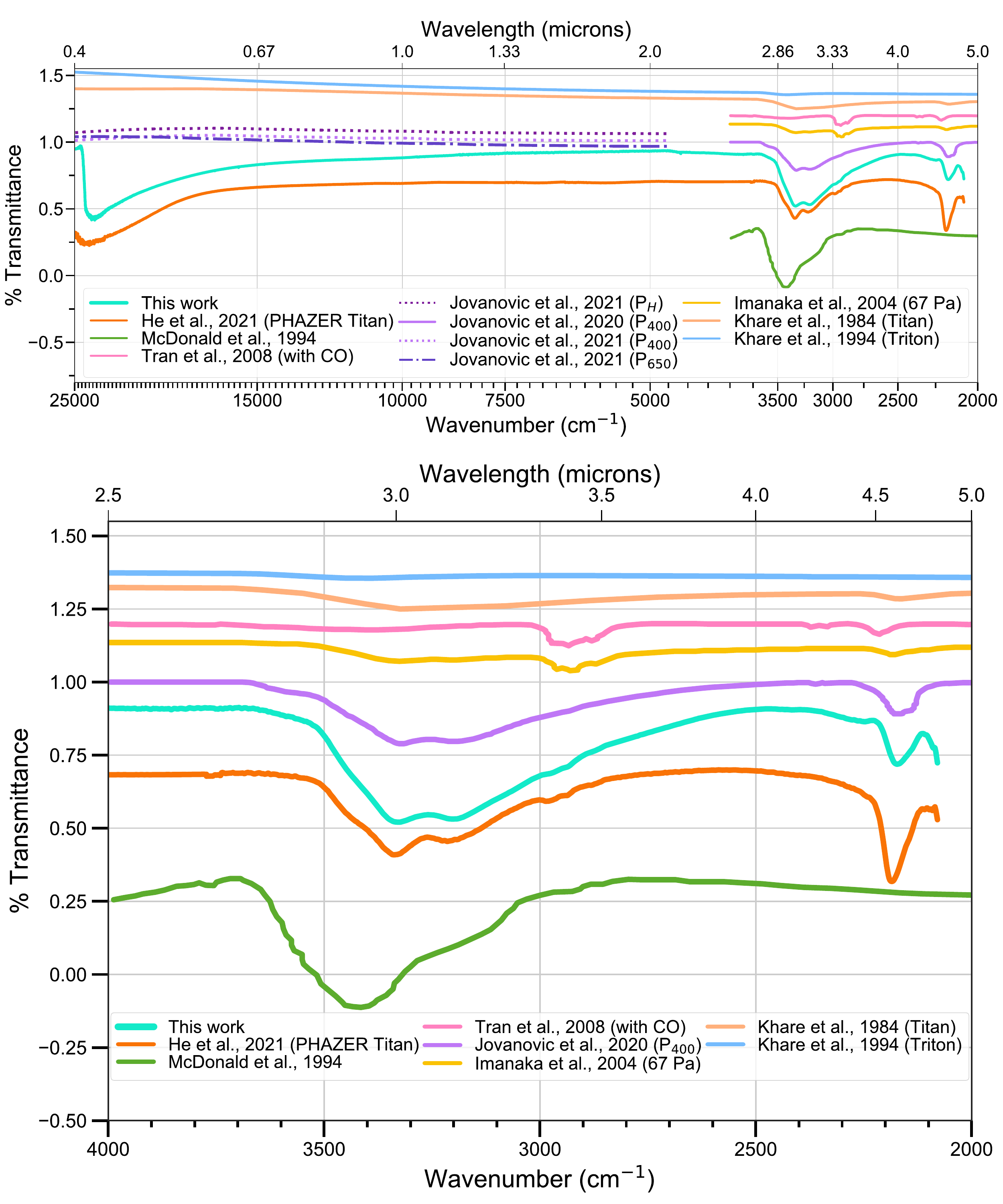}
\caption{TOP: Transmittance spectra of various tholins produced from N$_2$-CH$_4$ mixtures. The full details of each experiment can be found in Table \ref{table:exp_cond}. Some features are present in all tholin. In the Triton tholin of this work, we attribute some of the broadening and additional absorption between 3500 and 2500 cm$^{-1}$ to be due uniquely to O-H bonds. These bonds likely result from CO $>$ CH$_4$ in the initial gas mixture. All spectra have been offset vertically for clarity. BOTTOM: As above with emphasis on the 4000 cm$^{-1}$ to 2000 cm$^{-1}$ region.} 
\label{fig:transcompare}
\end{figure*}

We present a comparison of our Triton tholin transmittance spectra, PHAZER Titan transmittance spectra \citep{He2021titanftir}, and the other experimental data to which we compare our results in Figure \ref{fig:transcompare}. The PHAZER Titan tholin spectra has wider absorption at the blue visible edge (0.4 $\mu$m) compared to our Triton spectra, consistent with the slightly bluer cast of Triton's haze compared to Titan's, which is known from photometry \citep{hillier2021plutoradtransfer}. The most noticeable similarity amongst all the spectra is the broad feature between 3600 and 3200 cm$^{-1}$, which results from N-H bonds in the form of primary and secondary amines. The most overall similar spectra are the PHAZER Titan \citep{He2021titanftir} and Triton (this work) spectra, which is not surprising given the known effect of different experimental set-ups on resulting tholin properties \cite[e.g.,][]{cabletholinreview2012,brasse2015}. Our Triton tholin spectra have broad, wide features in the 3600 to 3200 cm$^{-1}$ region, which could also be due to the inclusion of oxygen in the chemical structure of the tholin, which could add O-H stretching from alcohols. This broadening extends to nearly 2500 cm$^{-1}$ and is apparent even in comparison to the PHAZER Titan data, which has absorption neither as wide nor as deep at these wavelengths. 

The second commonality to most spectra (except \citet{mcdonald1994titanandtritontholin} and \citet{khare1994triton} likely due to their very low starting methane mixing ratios) is a narrower feature at $\sim$ 2200 cm$^{-1}$ (4.5 $\mu$m). This absorption is likely from some combination of C$\equiv$C (alkyne) or C$\equiv$N (nitrile) bonds. While our Triton spectrum and \citet{JovanovicPlutoOrbitrap}'s Pluto spectrum have similarly wide and deep features here, the PHAZER Titan spectrum \citep{He2021titanftir} is markedly narrower and deeper at this feature, suggesting a different balance of alkynes to nitriles than that of the Pluto and Triton tholin.

Our Triton tholin, PHAZER Titan tholin \citep{He2021titanftir}, \citet{imanaka2004} tholin, and \citet{Tran2008} tholin all have features at $\sim$3000 cm$^{-1}$, which are indicative of C-H bonds in the form of alkenes or alkanes. These features are notably absent in the spectra of \citet{JovanovicPlutoOrbitrap}. \citet{He2021titanftir}, \citet{imanaka2004}, and \citet{Tran2008} all produced tholin from much higher methane mixing ratios (or in the case of Tran et. al, higher mixing ratios of other hydrocarbon species) than either \citet{JovanovicPlutoOrbitrap} or our Triton tholin. This suggests the additional absorption we see in the Triton tholin spectra in this region ($\sim$3000 to 2500 cm$^{-1}$) could derive from additional O-H bonds from alcohols and carboxylic acids that are not present in the Pluto composition tholin of \citet{JovanovicPlutoOrbitrap} because of the smaller amount of CO in their starting gas mixture. In a follow-up study, \citet{jovanovic2021opticalconstants} measured optical constants for Pluto analogue aerosols as well, varying the CH$_4$ mixing ratio but holding CO steady. In effect, this explored the importance of CO as it competes with CH$_4$. In agreement with our results here, their optical constants results show that N- and O-bearing molecules increase and generate additional absorption in the visible and near-IR with lower methane mixing ratios. Their study extended into the UV as well, finding that this absorption due to N and O molecules also affects shorter wavelengths \citep{jovanovic2021opticalconstants}.

We also note that many of the spectra to which we compare have their strongest identifying features at slightly longer wavelengths and into the fingerprint region, from roughly 5 to 15 $\mu$m (1700 to 700 cm$^{-1}$). This region is highly complex because of the multitude of features found there, but it can also provide much needed context to the VIS-NIR wavelength spectra explored here, particularly regarding the oxygen-bonding environment of tholin or haze particles. Carbonyl (C=O) species including aldehydes, ketones, esters, amides, and carboxylic acids all have prominent peaks from 1800 to 1600 cm$^{-1}$ (5.55 to 6.25 $\mu$m), and in fact have been observed in other studies with relatively larger CO starting fractions \citep{Tran2008}. Additionally, a study which examined the effect of CO$_2$ in N$_2$-CH$_4$ gas mixtures also found significant enhancement in both N- and O-bearing functional groups as CO$_2$ increased \citep{gavilan2018organicaerosols}. This study examined both shorter (UV, down to 0.13 $\mu$m) and longer (MIR, 6-10 $\mu$m) wavelengths, reitering the importance of wide wavelength coverage in understanding increasingly oxidized aerosols.

\begin{table}[hbtp!]
\caption{Summary of Experimental Conditions of the Tholin Spectra Presented in Figure \ref{fig:transcompare}.}
\begin{tabular}{llllll}
\toprule
\multicolumn{1}{l}{\textbf{Reference/Body}} & \multicolumn{1}{l}{\textbf{Gas mixture}} & \multicolumn{1}{l}{\textbf{Temperature}} & \multicolumn{1}{l}{\textbf{Pressure}} & \multicolumn{2}{l}{\textbf{Energy Source}} \\ 
\midrule
\begin{tabular}[c]{@{}l@{}}This work \\ \textit{Triton}\end{tabular}& \begin{tabular}[c]{@{}l@{}}99.3\% N$_2$\\ 0.5\% CO\\ 0.2\% CH$_4$\end{tabular} & 90 K & $\sim$1 mbar & \multicolumn{2}{l}{AC plasma discharge} \\
\addlinespace
\begin{tabular}[c]{@{}l@{}}He et al., 2021 \\ \textit{Titan} \end{tabular} & \begin{tabular}[c]{@{}l@{}}95\% N$_2$\\ 5\% CH$_4$\end{tabular} & 90 K & $\sim$1 mbar & \multicolumn{2}{l}{AC plasma  discharge} \\
\addlinespace
\begin{tabular}[c]{@{}l@{}}McDonald et al., 1994 \\ \textit{Triton} \end{tabular}& \begin{tabular}[c]{@{}l@{}}99.9\% N$_2$\\ 0.1\% CH$_4$\end{tabular} & \begin{tabular}[c]{@{}l@{}}room temperature\\ (294 K)\end{tabular} & $\sim$1 mbar & \multicolumn{2}{l}{RF ICP discharge} \\
\addlinespace
\begin{tabular}[c]{@{}l@{}}Tran et al., 2008 \\ \textit{Titan} \end{tabular}& \begin{tabular}[c]{@{}l@{}}98\% N$_2$\\ 1.8\% CH$_4$\\ 0.2\% H$_2$\\ 400 ppm C$_2$H$_2$\\ 300 ppm C$_2$H$_4$\\ 20 ppm HC$_3$N\\ 0.3\% CO\end{tabular} & 297 K & 900 mbar & \multicolumn{2}{l}{Mercury lamp} \\
\addlinespace
\begin{tabular}[c]{@{}l@{}}Jovanovi{\'c} et al., 2021 \\ \textit{Pluto, H ($<$350 km)} \end{tabular}& \begin{tabular}[c]{@{}l@{}}99.5\% N$_2$\\ 0.5\% CH$_4$\\ 500 ppm CO\end{tabular} & \begin{tabular}[c]{@{}l@{}}room temperature\\ (294 K)\end{tabular} & $\sim$1 mbar & \multicolumn{2}{l}{RF CCP discharge} \\
\addlinespace
\begin{tabular}[c]{@{}l@{}}Jovanovi{\'c} et al., 2020, 2021 \\ \textit{Pluto, 400 km} \end{tabular}& \begin{tabular}[c]{@{}l@{}}99\% N$_2$\\ 1\% CH$_4$\\ 500 ppm CO\end{tabular} & \begin{tabular}[c]{@{}l@{}}room temperature\\ (294 K)\end{tabular} & $\sim$1 mbar & \multicolumn{2}{l}{RF CCP discharge} \\
\addlinespace
\begin{tabular}[c]{@{}l@{}}Jovanovi{\'c} et al., 2021 \\ \textit{Pluto, 650 km} \end{tabular}& \begin{tabular}[c]{@{}l@{}}95\% N$_2$\\ 5\% CH$_4$\\ 500 ppm CO\end{tabular} & \begin{tabular}[c]{@{}l@{}}room temperature\\ (294 K)\end{tabular} & $\sim$1 mbar & \multicolumn{2}{l}{RF CCP discharge} \\
\addlinespace
\begin{tabular}[c]{@{}l@{}}Imanaka et al., 2004 \\ \textit{Titan} \end{tabular} & \begin{tabular}[c]{@{}l@{}}90\% N$_2$\\ 10\% CH$_4$\end{tabular} & \begin{tabular}[c]{@{}l@{}}room temperature\\ (294 K)\end{tabular} & 0.67 mbar & \multicolumn{2}{l}{RF ICP discharge} \\
\addlinespace
\begin{tabular}[c]{@{}l@{}} Khare et al., 1984 \\ \textit{Titan} \end{tabular}& \begin{tabular}[c]{@{}l@{}}90\% N$_2$\\ 10\% CH$_4$\end{tabular} & \begin{tabular}[c]{@{}l@{}}room temperature\\ (294 K)\end{tabular} & 0.2 mbar & \multicolumn{2}{l}{DC plasma discharge} \\ 
\addlinespace
\begin{tabular}[c]{@{}l@{}} Khare et al., 1994 \\ \textit{Triton} \end{tabular} & \begin{tabular}[c]{@{}l@{}}99.9\% N$_2$\\ 0.1\% CH$_4$\end{tabular} & \begin{tabular}[c]{@{}l@{}}room temperature\\ (294 K)\end{tabular} & 0.86 mbar & \multicolumn{2}{l}{DC plasma discharge}\\
\bottomrule
\end{tabular}
\\ \textit{Note.} AC: Alternating current. RF ICP: Radio frequency inductively coupled plasma. RF CCP: Radio frequency capacitively coupled plasma. DC: Direct current.
\label{table:exp_cond}
\end{table}
\newpage
\section{Discussion} \label{sec:discussion}

\subsection{Comparison to Existing Observations of Triton}

\subsubsection{Voyager 2 Atmospheric Observations}
\textit{In situ} observations of Triton's atmosphere are mainly limited to visible and UV photometry rather than spectra, preventing a true direct comparison of our experimental tholin results. Voyager 2 imaging and visible photometry observations of Triton's atmosphere suggest optically thin hazes \citep{smith1989}, with scattering optical depths between 0.001–0.01 that increase with shorter wavelengths \citep{pollack1990,ragesandpollack1992} and particle sizes of around 100-200 nm \citep{pollack1990,ragesandpollack1992}. UV occultation data from Voyager 2 suggests significantly higher (0.024) scattering optical depths, reiterating the wavelength dependent nature of the haze scattering and suggesting even smaller particles (30 nm) in the Rayleigh regime \citep{herbert1991}. At visible wavelengths, the haze is strongly forward scattering and requires particles larger than 200 nm \citep{hillier1991}. To meet these observational signatures of higher scattering in the blue (small particles) and large forward scattering (large particles) simultaneously, the hazes have been suggested to be fractal aggregates \citep{lavvas2021,ohno2021triton}. Aggregates also explain the haze properties of both Titan \citep{tomasko2008} and Pluto \citep{gladstone2016,cheng2017,gao2017,lavvas2021}, though the fractal dimension and growth rate is not settled \citep{Kutsop_2021}. This range of particle sizes well fits with those we produce in the PHAZER chamber \cite[e.g.,][]{he2017co,He2018}, though these particle sizes may in part result from the physical size of the chamber itself, which inhibits further growth. Material characterization of PHAZER Titan tholin particles suggests, however, that the necessary particle size and shape to match observations is easily achieved through coagulation \citep{yu2017}. More generally, tholin particles produced in a variety of setups seem to easily coagulate and form aggregates \citep{cabletholinreview2012}.

\subsubsection{Seasonal Changes from Space- and Ground-Based Observations}

Ground-based observations were responsible for the initial confirmed detection of CO in Triton's atmosphere \citep{lellouch2010}. However, earlier Hubble Space Telescope (HST) data using the Space Telescope Imaging Spectrograph (STIS) also provided upper limits on CO that differed from the initial Voyager 2 data and suggested strong surface absorption at UV wavelengths, which could be attributable to photochemical products from the atmosphere settling to the surface \citep{sternhstatmosphere1995,krasnopolsky1995tritonphotochem}. Both the ground- and space-based data indicate that CO levels in the atmosphere can change over seasonal timescales. Further HST STIS observations showed that Triton's UV albedo brightened between the Voyager 2 flyby and 1999, evidence of ongoing seasonal cycling in the form of volatile transport between the surface and atmosphere or plume activity
\citep{young2001hsttritonobs}. Later HST and ground-based observations continued to find surface reflectivity changes in the UV, as well as in visible and NIR wavelengths which probe methane abundances, suggestive of continued seasonal volatile transport on sub-decadal timescales \citep{bauer2010seasonalchanges,buratti2011}. Additionally, both thermal and pressure changes have been observed from ground-based solar occultations \citep{olkin1997,Elliot1998}, demonstrating that the atmosphere undergoes substantial interaction with the surface. The surface is predominately N$_2$ ice, with contributions from diluted CO ice \citep{merlin2018tritonicemeasurements}. This CO ice and N$_2$ ice experience greater longitudinal and temporal variation than does CH$_4$ ice present on the surface \citep{grundy2002tritonsurfacespectra,grundy2010}, including over diurnal timescales \citep{holler2016tritonice}.

Given our results regarding the importance of the exact CO/CH$_4$ mixing ratio in determining both the nitrogenation and oxidation of haze particles, these observations suggest that large shifts can happen in the dominant haze chemistry over Triton's seasonal cycle, as may also be occuring on Pluto \citep{bertrandandforget2017}. In their study of Pluto tholin optical constants, \citet{jovanovic2021opticalconstants} found that the real refractive index \textit{n} increased when the relative CO mixing ratio in the gas increased, and oxidation of the aerosols promoted both UV and visible absorption. Therefore, our experimental results combined with past observations suggests a complicated set of processes likely controls Triton's atmospheric and surface properties. Ices sublimate, change the overall atmospheric abundances, shift haze chemistry, and these hazes settle or condense out onto the surface, whereupon both ices and hazes can undergo additional alteration by cosmic rays and UV bombardment before potentially cycling back to the atmosphere again.

\subsection{Haze Formation and Ice Condensation}
\label{sec:icecondensation}

In light of this active surface-atmospheric interaction, a growing body of literature suggests that ice condensation, be it of photochemical products directly or with them as condensate nuclei for other ices \citep{luspaykuti2017,ohno2021triton}, could play a large  role in Triton's atmosphere. In fact, \citet{lavvas2021} argues that Titan, Pluto, and Triton lie along a continuum of haze formation outcomes, with Titan at the ``molecular growth'' end and Triton on the ``condensate growth'' end, with Pluto in the middle. These results depend on models which simulate organic ice formation \citep{lavvas2021}, but do not consider the implications of how photochemical haze formation may change due to the presence of CO, which experimental studies have shown can be an important driver in haze formation and growth \citep{horstandtolbert2014,he2017co,He2018,horst2018production,he2020}. While our experiment simulates an atmosphere with CO $>$ CH$_4$, as is present in Triton's atmosphere, we have increased their relative amounts compared to the background N$_2$ in order to increase our tholin yield within a shorter timeframe. Our experiment used 0.5\% CO and 0.2\% CH$_4$ while \citet{lellouch2010} calculated upper limits of 0.18\% CO and 0.034\% CH$_4$ in Triton's atmosphere.

While it may be tempting to interpret this as evidence for condensation processes over photochemical processes for the production of Triton's haze, many unknowns still exist, precluding a clear answer for the source of Triton's haze as of yet. Lower ratios (as is more reflective of the absolute abundances on Triton) can still generate substantial solid material photochemically. Previous work on N$_2$-CH$_4$ atmospheres has shown that, even with very low (0.1\%-0.01\%) methane mixing ratios, many solid particles are still generated \citep{trainer2004}, up to the same order in number and density of solids produced under more Titan-like methane mixing ratios \citep{horstandtolbert2013}. As discussed in Section \ref{sec:productionrates}, the role of the carbon source in haze production is not settled either, with varying amounts of CO and CH$_4$ not decisively showing higher yields across different experimental set-ups. While \citet{lavvas2021} argues that condensation of photochemical products occurs more quickly on Triton than molecular growth of solids can occur, these results rely on extrapolated nucleation and condensation rates for the materials in question. Our results, along with other laboratory measurements (see Section \ref{sec:productionrates}) suggests further study is warranted into the timescales of both processes for Triton, as well as on Pluto.

Future experimental work to simulate the uppermost reaches of Triton’s atmosphere where haze formation initiates, without CH$_4$, could help disentangle the relative roles of carbon monoxide versus methane in haze formation. Such work could isolate not only the relative importance of each in the chemical composition of the ultimate haze material, but also whether photochemical haze or ice condensates are more favorable for production of the haze observed in Triton’s atmosphere. Moreover, UV lamp and plasma produced tholin are known to sometimes differ in composition and spectra, hinting at differing mechanisms for haze initiation in the ionosphere and the lower atmosphere \citep{trainer2006,cabletholinreview2012}, which could also be investigated by future work with respect to Triton studies.

Prior comparisons arguing for ice condensates over photochemical products have also primarily used the optical constants of \citet{khare1984} Titan tholin \citep[e.g.,][]{ohno2021triton}. In the case of Pluto, more recent optical constants of purely photochemical Pluto tholin (from CO-containing mixtures) provide better agreements to existing New Horizons data of Pluto's surface \citep{jovanovic2021opticalconstants}, but there may still be a substantial contribution from condensed ices onto the hazes (see, e.g., \citealt{FAYOLLE2021114574}). Moreover, regardless of whether the particles are condensed ices or photochemical hazes that have settled to the surface, UV photons or cosmic ray bombardment could alter either set of particles, resulting in similar optical properties \citep[e.g.,][]{gudipati2013}. Our results, which show increased nitrogenation and oxidation even over that observed by \citet{JovanovicPlutoOrbitrap,jovanovic2021opticalconstants}, may further resolve observational discrepancies for Triton that do not necessarily require a substantial ice condensate component, or which could alter the behavior of hazes as potential condensate nuclei.

Further radiative transfer modeling has suggested that perhaps the redder Pluto and Titan hazes are more similar to each other in comparison to Triton, which is bluer. Such differences could be either chemical, perhaps due to higher CO/CH$_4$ mixing ratios as we have shown here, or physical, potentially due to the increased contribution from ice condensates in Triton's atmosphere \citep{hillier2021plutoradtransfer}. The question of whether Triton's (and Pluto's) hazes are solids produced from condensation or molecular growth is far from settled. To pinpoint the cause of the similarities and differences between the hazes of these worlds, further measurements to obtain optical constants of the spectra we provide here, laboratory measurements of the formation and properties of heterogeneous haze-ice condensate particles -- across multiple haze chemistries, and further modeling studies are all required.

\subsection{Comparison to Other Experimental Results}

We demonstrate above that the formulas for a variety of astrobiologically interesting molecules (amino acids, nucleobases, and a simple sugar) are present in our tholin, and subsequent sedimentation of this putative haze material likely coats the icy surface of Triton, as also likely occurs on Pluto \citep{grundy2018plutohazesurface,protopapa2020}. However, amino acids may be subject to further photolysis on icy surfaces, suggesting that they are rapidly destroyed unless this material quickly reaches the subsurface \citep{johnson2012tritoniceaminoacids}, though Triton's $\sim$14 $\mu$bar atmosphere attenuates more UV radiation than does Europa (with an atmosphere of 1 picobar) which \citet{johnson2012tritoniceaminoacids}'s experiment simulated. Complementing our atmospheric study, Triton surface ice chemistry by photolysis can also produce chemically complex material from N$_2$-CH$_4$-CO ice mixtures \citep{mooreandhudson2003,hodyss2011tritonicephotolysis}. The products of these surface ice experiments with CO/CH$_4$ $\sim$1 include many of the molecular species and spectral features we identify here, including carboxylic acids, alcohols, ketones, aldehydes, amines, and nitriles in addition to having similar nitrogen, carbon, and oxygen abundances overall \citep{materese2014,materese2015}. Additionally, potential interactions of these ices and tholin-like materials could further react with water ice deposits present on Triton's surface \citep{cruikshank2000}  to generate molecules of additional prebiotic complexity and interest \citep{cruikshank2019prebioticpluto}. Given the seasonal cycling through sublimation and condensation of volatile ices combined with the products of photochemistry both in the atmosphere and on the surface, Triton holds great interest for future study. 

\subsection{Future Triton Missions}

While ground- and space-based observations have furthered our understanding of Triton in the years since the Voyager 2 flyby \cite[e.g.,][]{sternhstatmosphere1995,young2001hsttritonobs,bauer2010seasonalchanges,lellouch2010,stansberry2015}, \textit{in situ} missions would dramatically improve our knowledge of the atmosphere, surface, and interior processes of the moon \citep{christophe2012,masters2014tritonaspartofsolarsystem,hofstadter2019,fletcher2020}. Several proposed missions in various states of development would visit Triton, with complementary goals. Trident, a Discovery class mission that was downselected as a finalist, though not ultimately chosen for flight in the Discovery 15 and 16 competition, would perform a single flyby of Triton \citep{prockter2019trident}. The Neptune Odyssey mission concept, a Flagship class orbiter and probe under study for the 2023 Planetary Science and Astrobiology Decadal Survey, would orbit the Neptune-Triton system with a four year prime mission and would perform $\sim$monthly flybys of Triton itself \citep{rymer2020}. Triton Hopper is a mission concept under study by the NASA Innovative Advanced Concepts (NIAC) program, which would act as a lander capable of producing its own propellant from surface ices in order to perform short flights across the surface \citep{Oleson2018,Landis2019MissionsTT}.

Trident's major scientific goals would be to confirm the presence of a subsurface ocean and infer whether it interacts with the surface, sample the ionosphere, and perform repeated surface imaging to characterize its composition and geology \citep{prockter2019trident}. To achieve these goals, Trident would be equipped with, among other instruments, a plasma spectrometer and a high-resolution infrared spectrometer with spectral range up to 5 $\mu$m. In the context of the study we have performed here, the Trident mission could clearly benefit our understanding of ionospheric processes driving haze formation by providing better constraints on the ion energies of the upper atmosphere. On Titan, for example, haze formation in the upper atmosphere is more strongly influenced by the Saturnian magnetosphere and solar EUV while longwave solar UV photons primarily dominate haze formation processes below 500 km \citep{lavvas2008}. In laboratory measurements, plasma or spark discharge energy sources compared to UV lamps produce tholin which differs in composition and observable spectra \citep{cabletholinreview2012}, illustrating this dual haze formation process. A better understanding of the Triton ionosphere could similarly help guide future laboratory experiments and modeling efforts to describe the haze formation process for Triton specifically. Trident's infrared spectrometer would have the same range as the spectral measurements we have performed in this study (out to 5 $\mu$m) and would clearly advance our understanding of Triton's surface and atmosphere-surface interactions \citep{prockter2019trident}. However, as we have shown above, the presence of CO as both a surface ice and and as a minor atmospheric component should induce significant oxygen chemistry. Such chemistry could be best probed with measurements that encompass carbonyl groups out past 6 $\mu$m, which could be taken into account with future development of the Trident concept.

The Triton Hopper concept study explored the ability of its design to generate propellant from the primarily nitrogen-ice surface of the moon, but also considered a wide instrument package to enable scientific characterization of Triton's surface and atmosphere \citep{Oleson2018,Landis2019MissionsTT}. These instruments include a quadrupole mass spectrometer and a gas chromatograph (based on SAM, Sample Analysis at Mars) a V/UV/NIR spectrometer, a meteorological package, and an X-ray spectrometer. In context of increased oxidation of sedimented haze materials on the surface suggested by our results, the ability of the Hopper concept to use surface materials as propellant may be impacted. Recent efforts to use Pluto-like tholin to match the New Horizons surface observations of dark reddish material suggest highly porous structures of surface ice mixed with aerosols \citep{FAYOLLE2021114574}. If similar porous ice-aerosol structure is present on Triton, or if aerosol is entrained with the surface ice or snow, the oxidized refractory organic materials that make up the haze could prevent efficient intake of N$_2$ to melt for propellant.

Neptune Odyssey, as a Flagship class concept, would be equipped with an extensive suite of instrumentation to enable its wide-ranging science goals, \textit{{\`a} la} a Cassini for the Neptune system \citep{rymer2020}. In terms of Triton science, Odyssey would investigate whether Triton is an ocean world, the source of its plumes, and broad scale compositional and dynamical processes of the atmosphere, surface, and interior. Relevant to our laboratory study here, this mission would carry imaging spectrometers in the UV and VIS-NIR, an ion and neutral mass spectrometer, a thermal plasma spectrometer, and an energetic charged particle detector. Such a mission would enable not only significant comparison to the Saturn-Titan system but also a far deeper understanding of Triton and Neptune than we currently possess. Nevertheless, the concept study also only studied a VIS-NIR spectrometer with range up to 5 $\mu$m and an ion and neutral mass spectrometer with range up to 100 amu \citep{rymer2020}. Here we have shown that significantly large and complex molecules with significant oxygen incorporation can be produced under Triton-like atmosphere conditions, which could be best explored with spectroscopy beyond 5 $\mu$m and mass spectrometry up to and beyond 450 amu. Lessons learned about the complex atmospheric chemistry of Titan and its unveiling (and the new questions uncovered) with Cassini should inform future mission development to explore and understand the correspondingly complex chemistry of Triton's atmosphere and surface.

\section{Conclusion} \label{sec:conclusion}

We simulated haze formation in Triton's atmosphere using the PHAZER chamber and apparatus with a starting gas mixture of 0.5\% CO and 0.2\% CH$_4$ in N$_2$ at 90 K. We then measured the production rate, composition, and spectra of the haze analogues produced with combustion analysis, very high resolution mass spectrometry, and transmittance and reflectance spectroscopy. We find that:

\begin{enumerate}
    \item Oxygen is incorporated into the elemental composition of the solid tholin particles at approximately 10\% by mass despite its inclusion in the form of CO in the original gas mixture at just 0.5\%.
    \item When taking our Triton results and comparing them to previous PHAZER N$_2$-CH$_4$-CO experiments, the increase of CO over that of CH$_4$ in the original gas mixture shifts the elemental composition away from carbon and toward a more nitrogen-rich structure, though more CH$_4$ may generate larger absolute amounts of solid.
    \item From very high resolution mass spectrometry measurements, we detect, as in previous studies, molecular formulas consistent with all 5 biological nucleotide bases, one non-biological nucleotide base, and several amino acid derivatives. Additionally, we observe the formula for glyceraldehyde, the simplest monosaccharide, for the first time from a N$_2$-CH$_4$-CO atmospheric experiment.
    \item Transmission and reflectance spectra of the Triton tholin produce features attributable to O-H, N-H, C-H, C$\equiv$N, and C$\equiv$C bonding. Most of these are also seen in spectra of similar tholin experimental data, but we also observe deeper and broader features, potentially attributable to oxygen bonds.
    \item To take full advantage of the chemistry we observe in our laboratory setting, future Triton missions should carry instrumentation capable of probing high molecular weight compounds, such as mass spectrometers with high mass range ($\geq$450 amu), and carbon-oxygen bonds, such as NIR spectrometers with spectral range out to at least 6.5 $\mu$m.
    \item The exact CO/CH$_4$ mixing ratio in N$_2$ atmospheres can dramatically affect the resulting haze chemistry and production rate. Since both Triton and Pluto undergo substantial atmospheric changes through sublimation of surface ices, their haze chemistry may also experience seasonal dependence.
\end{enumerate}

Given our results along with the existing body of literature for Titan and Pluto, additional study is clearly motivated into the exact chemical pathways for haze formation between these three similar, yet distinct planetary bodies. The nature of this haze chemistry can affect the prebiotic inventories of these worlds, their climates and radiative balance, and seasonal cycling between their atmospheres and surfaces. As three worlds with N$_2$-CH$_4$-CO atmospheres under different energetic regimes, Triton, Titan, and Pluto are themselves a fruitful laboratory for understanding carbon monoxide's dramatic influence on atmospheric chemistry.

\newpage
\appendix

\section{Complex Refractive Indices of Khare et al. 1994}
\label{sec:appendix}


\begin{longtable}{@{}lll@{}}
\toprule
\multicolumn{1}{c}{\textbf{Wavelength (microns)}} & \multicolumn{2}{c}{\textbf{Complex index of refraction}} \\* \midrule
\endfirsthead
\endhead
\bottomrule
\endfoot
\endlastfoot
 & \textit{Real index (n)} & \textit{Imaginary index (k)} \\
0.04959 & 0.86617 & 0.48009 \\
0.04999 & 0.87036 & 0.48617 \\
0.0504 & 0.87473 & 0.49233 \\
0.05081 & 0.87931 & 0.49861 \\
0.05123 & 0.88408 & 0.50499 \\
0.05166 & 0.88905 & 0.51148 \\
0.05209 & 0.89423 & 0.51807 \\
0.05253 & 0.89961 & 0.52479 \\
0.05298 & 0.9052 & 0.53161 \\
0.05344 & 0.91101 & 0.53856 \\
0.0539 & 0.91703 & 0.54561 \\
0.05438 & 0.92329 & 0.55279 \\
0.05486 & 0.92979 & 0.5601 \\
0.05535 & 0.93654 & 0.56755 \\
0.05585 & 0.94355 & 0.57513 \\
0.05635 & 0.95084 & 0.58285 \\
0.05687 & 0.95844 & 0.5907 \\
0.0574 & 0.96638 & 0.5987 \\
0.05793 & 0.97474 & 0.60686 \\
0.05848 & 0.9838 & 0.61515 \\
0.0588 & 0.98937 & 0.62 \\
0.05904 & 0.99352 & 0.6236 \\
0.05961 & 1.0031 & 0.63202 \\
0.06018 & 1.0124 & 0.64033 \\
0.06078 & 1.0216 & 0.64856 \\
0.06138 & 1.0309 & 0.65665 \\
0.06199 & 1.0404 & 0.66463 \\
0.06262 & 1.05 & 0.67247 \\
0.06325 & 1.0598 & 0.68016 \\
0.06391 & 1.0697 & 0.68767 \\
0.06457 & 1.0799 & 0.69501 \\
0.06525 & 1.0904 & 0.70215 \\
0.06595 & 1.1011 & 0.70907 \\
0.06666 & 1.1122 & 0.71576 \\
0.06738 & 1.1235 & 0.72218 \\
0.06812 & 1.1352 & 0.72834 \\
0.06888 & 1.1473 & 0.73419 \\
0.06965 & 1.1598 & 0.73972 \\
0.07044 & 1.1727 & 0.74489 \\
0.07125 & 1.1862 & 0.74968 \\
0.07208 & 1.2002 & 0.75405 \\
0.07293 & 1.2149 & 0.75798 \\
0.0738 & 1.2303 & 0.76143 \\
0.07469 & 1.2466 & 0.76436 \\
0.0756 & 1.2639 & 0.76673 \\
0.07653 & 1.2825 & 0.76849 \\
0.07749 & 1.303 & 0.76959 \\
0.07847 & 1.3275 & 0.76999 \\
0.0785 & 1.3284 & 0.76999 \\
0.07947 & 1.3551 & 0.76999 \\
0.08 & 1.3693 & 0.76998 \\
0.08051 & 1.3826 & 0.76886 \\
0.08157 & 1.4069 & 0.76294 \\
0.08265 & 1.4277 & 0.75498 \\
0.08377 & 1.447 & 0.74602 \\
0.08492 & 1.4651 & 0.73642 \\
0.0861 & 1.4825 & 0.72633 \\
0.08731 & 1.4994 & 0.7158 \\
0.08856 & 1.516 & 0.70487 \\
0.08984 & 1.5324 & 0.69356 \\
0.09116 & 1.5492 & 0.68186 \\
0.0925 & 1.5678 & 0.66996 \\
0.09252 & 1.5681 & 0.66975 \\
0.09392 & 1.5864 & 0.65592 \\
0.09537 & 1.6018 & 0.64162 \\
0.09686 & 1.6159 & 0.62687 \\
0.0984 & 1.6293 & 0.61166 \\
0.09998 & 1.6422 & 0.59597 \\
0.1016 & 1.6554 & 0.58 \\
0.10162 & 1.6556 & 0.5798 \\
0.10332 & 1.6683 & 0.56301 \\
0.10507 & 1.6791 & 0.54572 \\
0.10688 & 1.689 & 0.52785 \\
0.10875 & 1.6992 & 0.5094 \\
0.1097 & 1.7042 & 0.49999 \\
0.1107 & 1.7092 & 0.48935 \\
0.11271 & 1.7164 & 0.46748 \\
0.1148 & 1.721 & 0.44355 \\
0.1159 & 1.7214 & 0.42999 \\
0.11696 & 1.7214 & 0.41559 \\
0.1181 & 1.7184 & 0.40008 \\
0.11921 & 1.7143 & 0.38848 \\
0.1215 & 1.6999 & 0.37 \\
0.12155 & 1.6996 & 0.3698 \\
0.12398 & 1.6837 & 0.36075 \\
0.126 & 1.6713 & 0.35602 \\
0.12651 & 1.6685 & 0.35704 \\
0.12915 & 1.6585 & 0.36875 \\
0.1305 & 1.6598 & 0.3752 \\
0.13189 & 1.6615 & 0.38313 \\
0.13476 & 1.6846 & 0.40345 \\
0.1366 & 1.7114 & 0.4143 \\
0.13776 & 1.7275 & 0.4148 \\
0.14089 & 1.7588 & 0.37951 \\
0.14416 & 1.7658 & 0.31228 \\
0.1476 & 1.7544 & 0.25052 \\
0.1503 & 1.7315 & 0.24102 \\
0.1512 & 1.7264 & 0.248 \\
0.1532 & 1.7249 & 0.26535 \\
0.15498 & 1.7235 & 0.25909 \\
0.1559 & 1.7218 & 0.25492 \\
0.1585 & 1.7139 & 0.2095 \\
0.15895 & 1.7124 & 0.20955 \\
0.16313 & 1.6781 & 0.21051 \\
0.16754 & 1.6612 & 0.21258 \\
0.17219 & 1.6442 & 0.21626 \\
0.175 & 1.6347 & 0.2194 \\
0.17711 & 1.6275 & 0.22896 \\
0.18 & 1.6158 & 0.24359 \\
0.18232 & 1.6067 & 0.24648 \\
0.1845 & 1.6083 & 0.24932 \\
0.18785 & 1.6121 & 0.30502 \\
0.1879 & 1.6123 & 0.30535 \\
0.19372 & 1.6575 & 0.33576 \\
0.195 & 1.667 & 0.33758 \\
0.19997 & 1.7024 & 0.32769 \\
0.2028 & 1.7168 & 0.3217 \\
0.2049 & 1.7271 & 0.31721 \\
0.20663 & 1.7355 & 0.31433 \\
0.2097 & 1.7503 & 0.3094 \\
0.21376 & 1.7694 & 0.3018 \\
0.215 & 1.7751 & 0.2992 \\
0.22139 & 1.8033 & 0.2748 \\
0.2264 & 1.8195 & 0.251 \\
0.22959 & 1.827 & 0.2328 \\
0.2364 & 1.8329 & 0.19104 \\
0.23842 & 1.8336 & 0.18305 \\
0.2471 & 1.8295 & 0.1491 \\
0.24796 & 1.8289 & 0.14573 \\
0.25829 & 1.8142 & 0.1051 \\
0.2587 & 1.8135 & 0.1035 \\
0.261 & 1.809 & 0.094679 \\
0.26952 & 1.791 & 0.082132 \\
0.273 & 1.7853 & 0.078 \\
0.28177 & 1.7732 & 0.069208 \\
0.29519 & 1.76 & 0.057937 \\
0.3 & 1.756 & 0.05411 \\
0.30995 & 1.7482 & 0.046466 \\
0.324 & 1.7376 & 0.0369 \\
0.32626 & 1.736 & 0.03559 \\
0.34439 & 1.7239 & 0.026779 \\
0.36 & 1.7143 & 0.0209 \\
0.36465 & 1.7117 & 0.019356 \\
0.37 & 1.7088 & 0.017767 \\
0.38 & 1.7036 & 0.01573 \\
0.38744 & 1.7001 & 0.015037 \\
0.39 & 1.699 & 0.014736 \\
0.4 & 1.6949 & 0.012734 \\
0.41327 & 1.6901 & 0.010889 \\
0.42 & 1.6878 & 0.010059 \\
0.44 & 1.6817 & 0.0078597 \\
0.44279 & 1.6809 & 0.0075893 \\
0.46 & 1.6762 & 0.0061141 \\
0.47685 & 1.6721 & 0.0049604 \\
0.48 & 1.6714 & 0.004791 \\
0.5 & 1.6672 & 0.0041001 \\
0.51658 & 1.6641 & 0.0037892 \\
0.53 & 1.6619 & 0.003574 \\
0.55 & 1.6591 & 0.00328 \\
0.56355 & 1.6574 & 0.0031236 \\
0.57 & 1.6566 & 0.0030584 \\
0.59 & 1.6545 & 0.0028893 \\
0.6 & 1.6536 & 0.0028187 \\
0.61 & 1.6527 & 0.0027664 \\
0.6199 & 1.6518 & 0.0027253 \\
0.62 & 1.6518 & 0.0027249 \\
0.63 & 1.651 & 0.0026867 \\
0.64 & 1.6502 & 0.0026499 \\
0.65 & 1.6495 & 0.0026138 \\
0.66 & 1.6488 & 0.0025789 \\
0.67 & 1.6481 & 0.0025461 \\
0.68 & 1.6475 & 0.0025154 \\
0.68878 & 1.647 & 0.0024901 \\
0.69 & 1.6469 & 0.0024867 \\
0.7 & 1.6464 & 0.00246 \\
0.72 & 1.6454 & 0.0024134 \\
0.74 & 1.6445 & 0.0023746 \\
0.75 & 1.6441 & 0.0023575 \\
0.77 & 1.6433 & 0.0023314 \\
0.77488 & 1.6431 & 0.0023267 \\
0.79 & 1.6426 & 0.0023144 \\
0.8 & 1.6422 & 0.0022959 \\
0.82 & 1.6416 & 0.002197 \\
0.84 & 1.6411 & 0.0020675 \\
0.86 & 1.6405 & 0.0020144 \\
0.88 & 1.64 & 0.0019352 \\
0.88557 & 1.6399 & 0.0019106 \\
0.9 & 1.6396 & 0.0018452 \\
0.95 & 1.6385 & 0.0016617 \\
1 & 1.6376 & 0.0015323 \\
1.0332 & 1.637 & 0.0014503 \\
1.1 & 1.6361 & 0.001298 \\
1.2 & 1.635 & 0.0011223 \\
1.2398 & 1.6346 & 0.0010529 \\
1.3 & 1.6341 & 0.00094895 \\
1.4 & 1.6333 & 0.0008075 \\
1.5 & 1.6326 & 0.00067144 \\
1.5498 & 1.6323 & 0.00062089 \\
1.6 & 1.632 & 0.0005775 \\
1.7 & 1.6314 & 0.00050698 \\
1.8 & 1.6307 & 0.0004711 \\
1.9 & 1.6302 & 0.00040131 \\
2 & 1.6297 & 0.00028705 \\
2.0663 & 1.6299 & 0.00026788 \\
2.1 & 1.6304 & 0.00025236 \\
2.1014 & 1.6304 & 0.00025099 \\
2.1376 & 1.6299 & 0.00019661 \\
2.1751 & 1.6296 & 0.00012606 \\
2.2 & 1.6294 & 9.36E-05 \\
2.2139 & 1.6293 & 9.23E-05 \\
2.2542 & 1.6291 & 9.49E-05 \\
2.2959 & 1.6288 & 9.58E-05 \\
2.3 & 1.6288 & 9.58E-05 \\
2.3392 & 1.6286 & 9.54E-05 \\
2.3842 & 1.6283 & 9.44E-05 \\
2.4 & 1.6282 & 9.18E-05 \\
2.431 & 1.628 & 6.72E-05 \\
2.4796 & 1.6276 & 1.37E-05 \\
2.6379 & 1.6252 & 3.80E-06 \\
2.6455 & 1.625 & 1.48E-05 \\
2.6667 & 1.6242 & 0.00018093 \\
2.6952 & 1.6231 & 0.00041332 \\
2.7027 & 1.6228 & 0.00048627 \\
2.7397 & 1.6211 & 0.0018205 \\
2.7551 & 1.6205 & 0.0031149 \\
2.7778 & 1.6202 & 0.0051794 \\
2.8169 & 1.6207 & 0.0088768 \\
2.8177 & 1.6207 & 0.0089462 \\
2.8571 & 1.6232 & 0.011993 \\
2.8833 & 1.6253 & 0.013656 \\
2.8986 & 1.6267 & 0.01434 \\
2.9412 & 1.6309 & 0.013891 \\
2.9519 & 1.6318 & 0.013396 \\
2.9851 & 1.6338 & 0.011394 \\
3.0239 & 1.6353 & 0.0086455 \\
3.0303 & 1.6354 & 0.0082556 \\
3.0769 & 1.6357 & 0.0064441 \\
3.0995 & 1.6357 & 0.0056691 \\
3.125 & 1.6356 & 0.0047953 \\
3.1746 & 1.6352 & 0.003073 \\
3.179 & 1.6352 & 0.0029589 \\
3.2258 & 1.6345 & 0.0020468 \\
3.2626 & 1.634 & 0.0014496 \\
3.2787 & 1.6338 & 0.0012418 \\
3.3333 & 1.6331 & 0.00084332 \\
3.3508 & 1.6329 & 0.00073703 \\
3.3898 & 1.6325 & 0.00053206 \\
3.4439 & 1.632 & 0.00030838 \\
3.4483 & 1.632 & 0.00029497 \\
3.5088 & 1.6315 & 0.00018484 \\
3.5423 & 1.6312 & 0.00015032 \\
3.5714 & 1.631 & 0.00012355 \\
3.6364 & 1.6307 & 6.89E-05 \\
3.6465 & 1.6306 & 6.23E-05 \\
3.7037 & 1.6303 & 3.46E-05 \\
3.757 & 1.6301 & 3.06E-05 \\
3.7736 & 1.63 & 2.83E-05 \\
3.8461 & 1.6297 & 7.01E-06 \\
3.8744 & 1.6296 & 3.15E-06 \\
4.065 & 1.6288 & 9.24E-06 \\
4.0816 & 1.6287 & 1.48E-05 \\
4.1169 & 1.6286 & 2.28E-05 \\
4.1327 & 1.6285 & 3.79E-05 \\
4.1459 & 1.6285 & 5.20E-05 \\
4.1667 & 1.6285 & 7.53E-05 \\
4.1806 & 1.6285 & 0.00010292 \\
4.2087 & 1.6284 & 0.00057176 \\
4.2319 & 1.6284 & 0.00093355 \\
4.2553 & 1.6284 & 0.00063071 \\
4.268 & 1.6284 & 0.00075411 \\
4.2752 & 1.6284 & 0.00072837 \\
4.2827 & 1.6284 & 0.00069512 \\
4.2937 & 1.6284 & 0.00063988 \\
4.3122 & 1.6284 & 0.00047647 \\
4.329 & 1.6284 & 0.00035105 \\
4.3478 & 1.6284 & 0.00025447 \\
4.3649 & 1.6283 & 0.00020347 \\
4.3764 & 1.6283 & 0.00020734 \\
4.4053 & 1.6283 & 0.00024646 \\
4.4189 & 1.6283 & 0.00026707 \\
4.4279 & 1.6283 & 0.00026953 \\
4.4287 & 1.6283 & 0.00027031 \\
4.4385 & 1.6283 & 0.00028135 \\
4.4444 & 1.6283 & 0.00028594 \\
4.4683 & 1.6282 & 0.0003004 \\
4.4763 & 1.6282 & 0.00031924 \\
4.5126 & 1.6281 & 0.00038237 \\
4.529 & 1.628 & 0.00039904 \\
4.5455 & 1.628 & 0.00041788 \\
4.562 & 1.628 & 0.00041695 \\
4.5788 & 1.6279 & 0.00042681 \\
4.5914 & 1.6279 & 0.00042622 \\
4.5919 & 1.6279 & 0.00042621 \\
4.6104 & 1.6279 & 0.00043211 \\
4.6211 & 1.6278 & 0.00042559 \\
4.6318 & 1.6278 & 0.0004171 \\
4.6512 & 1.6278 & 0.00040957 \\
4.6729 & 1.6278 & 0.00039339 \\
4.6904 & 1.6277 & 0.00039101 \\
4.7148 & 1.6277 & 0.00038963 \\
4.7259 & 1.6277 & 0.00038724 \\
4.7619 & 1.6276 & 0.00037006 \\
4.7685 & 1.6276 & 0.00036763 \\
4.7916 & 1.6276 & 0.00036111 \\
4.8263 & 1.6275 & 0.00035686 \\
4.8426 & 1.6275 & 0.00035122 \\
4.878 & 1.6274 & 0.000331 \\
4.8876 & 1.6274 & 0.0003271 \\
4.9116 & 1.6273 & 0.00032356 \\
4.931 & 1.6273 & 0.00030838 \\
4.953 & 1.6272 & 0.00030357 \\
4.9579 & 1.6272 & 0.00029346 \\
4.9592 & 1.6272 & 0.00029069 \\
4.9751 & 1.6272 & 0.00028162 \\
4.9826 & 1.6271 & 0.00027436 \\
4.985 & 1.6271 & 0.00027147 \\
5.0176 & 1.6271 & 0.0002642 \\
5.0302 & 1.627 & 0.0002509 \\
5.0454 & 1.627 & 0.00023347 \\
5.0607 & 1.627 & 0.00022653 \\
5.0839 & 1.6269 & 0.00021348 \\
5.1099 & 1.6269 & 0.00018989 \\
5.1282 & 1.6268 & 0.00018452 \\
5.1414 & 1.6268 & 0.00017904 \\
5.152 & 1.6267 & 0.00016075 \\
5.1658 & 1.6267 & 0.00015504 \\
5.1733 & 1.6267 & 0.00015271 \\
5.2111 & 1.6266 & 0.00013814 \\
5.2356 & 1.6265 & 0.00010829 \\
5.2632 & 1.6264 & 4.57E-06 \\
5.4764 & 1.6252 & 9.97E-06 \\
5.5157 & 1.6248 & 4.84E-05 \\
5.5371 & 1.6247 & 9.39E-05 \\
5.5556 & 1.6245 & 0.00010655 \\
5.6117 & 1.6241 & 0.00014623 \\
5.6338 & 1.6239 & 0.0001589 \\
5.6355 & 1.6239 & 0.00016004 \\
5.6818 & 1.6236 & 0.00023133 \\
5.7143 & 1.6234 & 0.00040635 \\
5.7504 & 1.6231 & 0.00077836 \\
5.7703 & 1.623 & 0.0010454 \\
5.7937 & 1.6228 & 0.001056 \\
5.8173 & 1.6227 & 0.0011696 \\
5.8275 & 1.6226 & 0.0012261 \\
5.8411 & 1.6225 & 0.0013033 \\
5.8823 & 1.6223 & 0.0022336 \\
5.9038 & 1.6223 & 0.0027969 \\
5.9277 & 1.6225 & 0.0035315 \\
5.9595 & 1.6228 & 0.0049669 \\
5.9844 & 1.6231 & 0.0065636 \\
5.9988 & 1.6233 & 0.007372 \\
6.0387 & 1.6239 & 0.0083311 \\
6.0606 & 1.6242 & 0.0090692 \\
6.0939 & 1.6246 & 0.010174 \\
6.1237 & 1.625 & 0.010441 \\
6.1652 & 1.6254 & 0.0093897 \\
6.199 & 1.6258 & 0.0074053 \\
6.2112 & 1.6259 & 0.0067416 \\
6.25 & 1.6263 & 0.0050224 \\
6.3211 & 1.627 & 0.003202 \\
6.3452 & 1.6272 & 0.0029247 \\
6.3816 & 1.6275 & 0.0025698 \\
6.4103 & 1.6277 & 0.0025929 \\
6.4267 & 1.6278 & 0.0024455 \\
6.4516 & 1.628 & 0.002453 \\
6.4641 & 1.6281 & 0.0024269 \\
6.4809 & 1.6281 & 0.0024937 \\
6.5062 & 1.6283 & 0.0022592 \\
6.5253 & 1.6283 & 0.0021635 \\
6.536 & 1.6283 & 0.0021386 \\
6.566 & 1.6282 & 0.0021588 \\
6.5789 & 1.6282 & 0.0021253 \\
6.605 & 1.628 & 0.0021887 \\
6.6269 & 1.6279 & 0.0022664 \\
6.6445 & 1.6278 & 0.0022082 \\
6.6667 & 1.6276 & 0.0022567 \\
6.6845 & 1.6275 & 0.0022392 \\
6.7069 & 1.6273 & 0.0021785 \\
6.7249 & 1.6272 & 0.0020343 \\
6.7568 & 1.627 & 0.0019761 \\
6.7797 & 1.6269 & 0.0019367 \\
6.8027 & 1.6268 & 0.0018156 \\
6.8259 & 1.6267 & 0.0017637 \\
6.8353 & 1.6266 & 0.0017266 \\
6.8493 & 1.6266 & 0.0016424 \\
6.8776 & 1.6264 & 0.0015658 \\
6.8878 & 1.6264 & 0.0015594 \\
6.8966 & 1.6264 & 0.0015542 \\
6.9348 & 1.6263 & 0.0014976 \\
6.9541 & 1.6262 & 0.0015125 \\
6.9735 & 1.6262 & 0.0014798 \\
7.0126 & 1.6261 & 0.0014863 \\
7.0423 & 1.626 & 0.0014961 \\
7.0622 & 1.626 & 0.00145 \\
7.0972 & 1.6259 & 0.0014613 \\
7.1429 & 1.6258 & 0.0013825 \\
7.2411 & 1.6256 & 0.0012282 \\
7.2929 & 1.6255 & 0.0011018 \\
7.3314 & 1.6254 & 0.0010048 \\
7.3855 & 1.6253 & 0.00086976 \\
7.4074 & 1.6252 & 0.00086151 \\
7.4516 & 1.6251 & 0.00081194 \\
7.4738 & 1.625 & 0.00077195 \\
7.5244 & 1.6249 & 0.00069306 \\
7.57 & 1.6247 & 0.00062427 \\
7.593 & 1.6247 & 0.00055246 \\
7.6923 & 1.6244 & 0.0004644 \\
7.7399 & 1.6242 & 0.00044747 \\
7.7488 & 1.6242 & 0.0004461 \\
7.8309 & 1.6239 & 0.00044524 \\
7.9681 & 1.6234 & 0.00042526 \\
8.1235 & 1.6229 & 0.00040333 \\
8.2169 & 1.6226 & 0.0003946 \\
8.2653 & 1.6224 & 0.00039497 \\
8.2988 & 1.6223 & 0.00039861 \\
8.3333 & 1.6222 & 0.00040807 \\
8.4104 & 1.622 & 0.00046919 \\
8.4818 & 1.6219 & 0.00068084 \\
8.5543 & 1.6217 & 0.0008507 \\
8.6059 & 1.6216 & 0.00095633 \\
8.6957 & 1.6214 & 0.0011349 \\
8.7489 & 1.6213 & 0.0011455 \\
8.8028 & 1.6211 & 0.00099961 \\
8.8557 & 1.621 & 0.00089411 \\
8.8731 & 1.621 & 0.00089171 \\
8.8968 & 1.6209 & 0.00097744 \\
8.9365 & 1.6208 & 0.0010451 \\
8.9928 & 1.6207 & 0.00086646 \\
9.0498 & 1.6205 & 0.00064804 \\
9.0909 & 1.6204 & 0.00054689 \\
9.1575 & 1.6203 & 0.00042919 \\
9.311 & 1.6199 & 0.00028792 \\
9.5369 & 1.6193 & 0.00025544 \\
9.5694 & 1.6192 & 0.00025166 \\
9.8912 & 1.6179 & 0.00022374 \\
10 & 1.6175 & 0.00018944 \\
10.204 & 1.6166 & 5.81E-05 \\
10.267 & 1.6163 & 2.22E-06 \\
10.73 & 1.6141 & 4.92E-06 \\
10.788 & 1.6138 & 0.00010207 \\
10.881 & 1.6133 & 0.00016606 \\
11.025 & 1.6126 & 0.00022157 \\
11.111 & 1.6122 & 0.0002574 \\
11.198 & 1.6118 & 0.00036696 \\
11.271 & 1.6114 & 0.00052602 \\
11.274 & 1.6114 & 0.00053403 \\
11.338 & 1.6111 & 0.00072421 \\
11.415 & 1.6107 & 0.00097889 \\
11.481 & 1.6103 & 0.00085287 \\
11.601 & 1.6097 & 0.00099804 \\
11.765 & 1.6089 & 0.0012961 \\
12.034 & 1.6075 & 0.0019181 \\
12.315 & 1.6061 & 0.0027455 \\
12.398 & 1.6057 & 0.0029207 \\
12.407 & 1.6057 & 0.0029374 \\
12.5 & 1.6052 & 0.0030883 \\
12.739 & 1.6042 & 0.0038279 \\
12.937 & 1.6033 & 0.0044139 \\
13.004 & 1.603 & 0.0045061 \\
13.038 & 1.6029 & 0.0045918 \\
13.051 & 1.6028 & 0.0046262 \\
13.333 & 1.6013 & 0.0054012 \\
13.459 & 1.6006 & 0.0058752 \\
13.624 & 1.5997 & 0.0063335 \\
13.717 & 1.5992 & 0.0069037 \\
13.776 & 1.5989 & 0.0071453 \\
14.045 & 1.5976 & 0.008045 \\
14.286 & 1.5965 & 0.0090467 \\
14.586 & 1.5953 & 0.010598 \\
14.859 & 1.5943 & 0.012258 \\
14.993 & 1.5938 & 0.013134 \\
15.06 & 1.5936 & 0.01295 \\
15.385 & 1.5925 & 0.014476 \\
15.498 & 1.5922 & 0.014993 \\
16.051 & 1.5908 & 0.017311 \\
16.26 & 1.5902 & 0.018032 \\
16.531 & 1.5895 & 0.019364 \\
16.667 & 1.5891 & 0.019994 \\
17.123 & 1.5879 & 0.021403 \\
17.391 & 1.5872 & 0.022192 \\
17.637 & 1.5866 & 0.022923 \\
17.711 & 1.5864 & 0.022985 \\
17.825 & 1.5859 & 0.023049 \\
18.018 & 1.585 & 0.023671 \\
18.182 & 1.5842 & 0.023889 \\
18.416 & 1.5828 & 0.024375 \\
18.484 & 1.5824 & 0.024511 \\
18.553 & 1.582 & 0.024524 \\
18.657 & 1.5814 & 0.024404 \\
18.832 & 1.5802 & 0.024658 \\
19.011 & 1.579 & 0.024725 \\
19.074 & 1.5786 & 0.024867 \\
19.231 & 1.5775 & 0.025309 \\
19.531 & 1.575 & 0.024931 \\
19.685 & 1.5737 & 0.02467 \\
20 & 1.5708 & 0.024901 \\
20.663 & 1.5654 & 0.033283 \\
22.222 & 1.5615 & 0.057638 \\
22.542 & 1.5614 & 0.060825 \\
24.796 & 1.567 & 0.073943 \\
25 & 1.5662 & 0.07471 \\
27.551 & 1.5511 & 0.074067 \\
28.571 & 1.5456 & 0.076355 \\
30.995 & 1.5342 & 0.12651 \\
33.333 & 1.5457 & 0.17464 \\
35.423 & 1.5629 & 0.20358 \\
40 & 1.6181 & 0.24374 \\
41.327 & 1.6329 & 0.25248 \\
49.592 & 1.703 & 0.30692 \\
61.99 & 1.758 & 0.38857 \\
82.653 & 1.8148 & 0.52467 \\
123.98 & 1.8878 & 0.79687 \\* \bottomrule
\caption{The complex refractive indices of Triton tholin from \citet{khare1994triton}.}
\label{table:khare94}
\end{longtable}

\newpage
\section{Blank Mass Spectra}
\label{sec:blanks}
\begin{figure*}[hbt!]
\centering
\includegraphics[angle=0,width=0.95\linewidth]{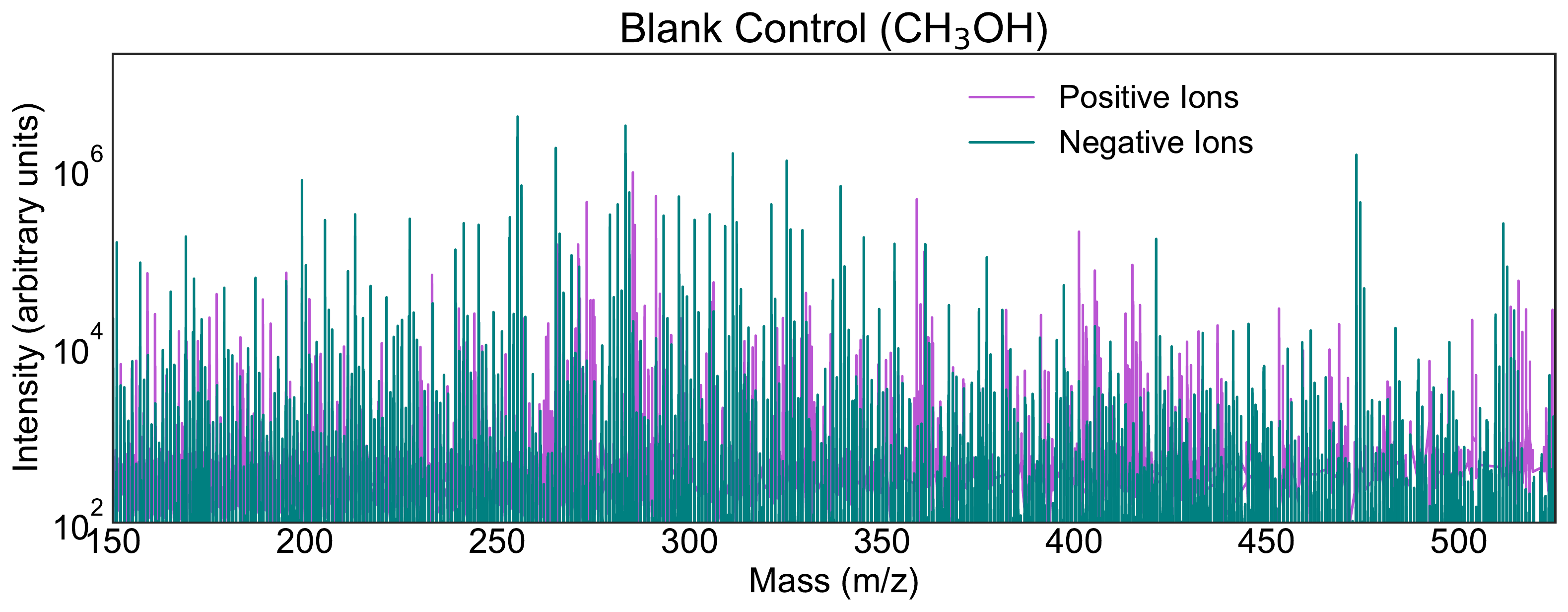}
\caption{Positive (magenta) and negative (teal) ion mode mass spectra of the blank solution of CH$_3$OH from m/z 150 to 525. The blank clearly lacks the structure of the actual sample (see Figure \ref{fig:mass_spec}). Peaks greater than 2$\times$10$^{5}$ present in the blank were removed from the Triton samples before peak assignments were made. These removed peaks constitute less than 0.03\% of all peaks.}
\label{fig:blank}
\end{figure*}

\begin{figure*}[hbt!]
\centering
\includegraphics[angle=0,width=0.95\linewidth]{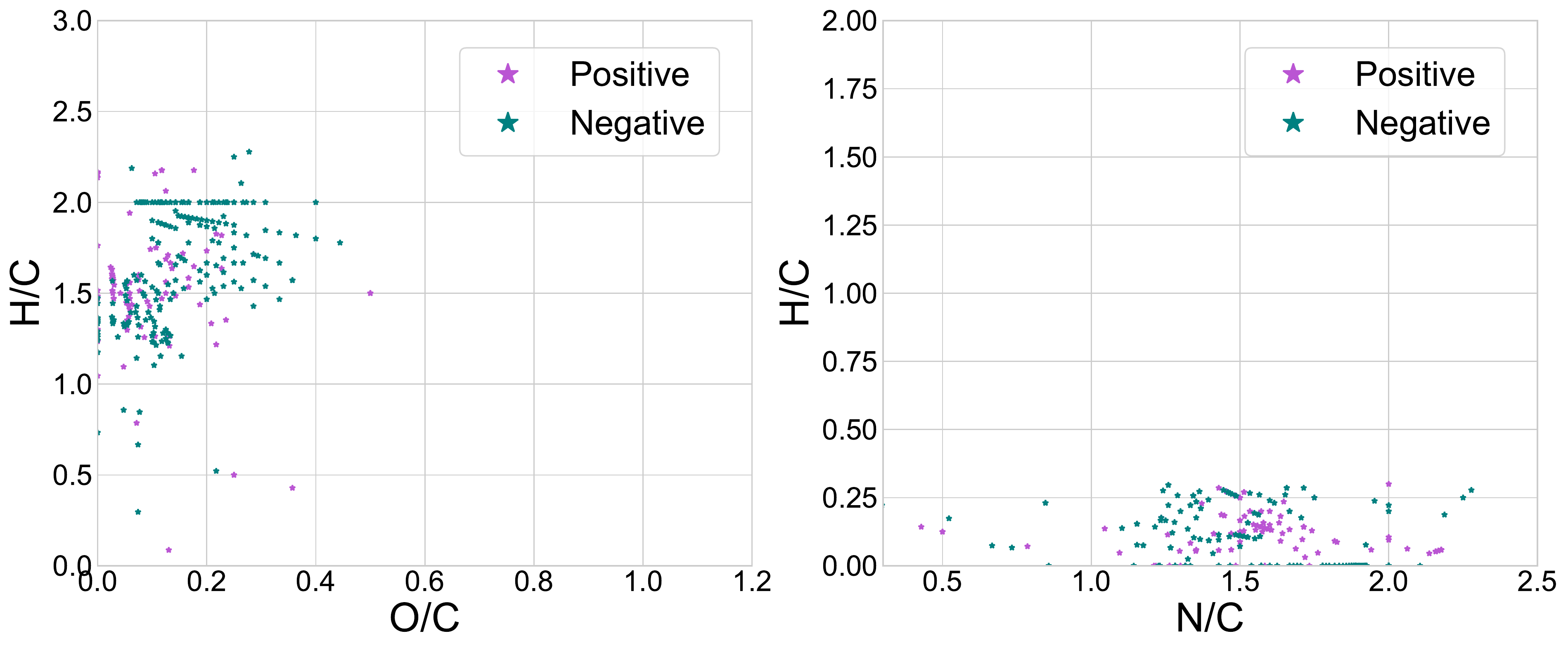}
\caption{Positive (magenta) and negative (teal) ion mode van Krevelen diagrams of the blank solvent control. Left: H/C vs O/C. Right: H/C vs N/C.}
\label{fig:vk_blank}
\end{figure*}

\newpage
\section{Fringe Correction}
\label{sec:appendix_fringes}

\begin{figure*}[hbt!]
\centering
\includegraphics[angle=0,width=0.95\linewidth]{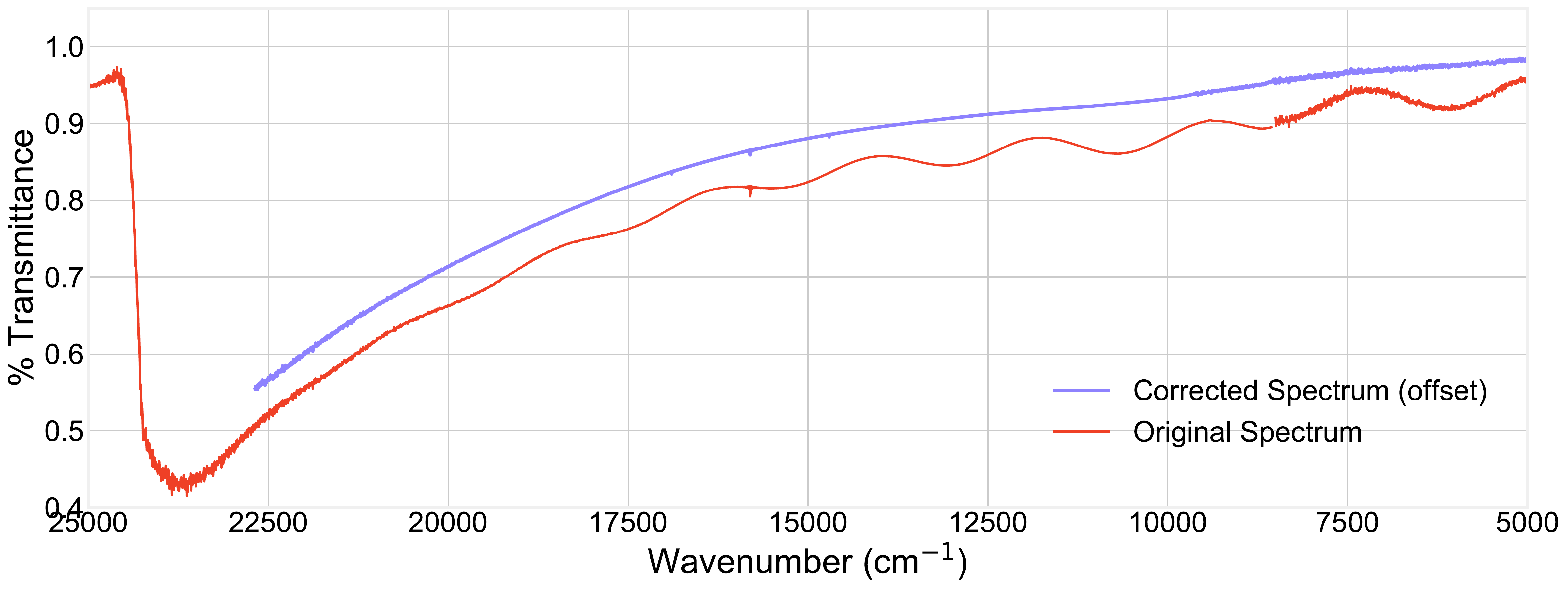}
\caption{Raw (red) and fringe-corrected transmittance spectrum of the Triton haze analogues.}
\label{fig:fringes}
\end{figure*}

\acknowledgments
S.E. Moran was supported by NASA Earth and Space Science Fellowship grant 80NSSC18K1109. S.E. Moran thanks R.M. Moran for enlightening discussions during development of the manuscript. This work is supported by the French National Research Agency in the framework of the Investissements d'Avenir program (ANR-15-IDEX-02), through the funding of the ``Origin of Life'' project of the Univ. Grenoble-Alpes and the French Space Agency (CNES) under their Exobiology and Solar System programs. C. Wolters acknowledges a PhD fellowship from CNES/ANR (ANR-16-CE29-0015 2016-2021). We also thank Darrell Strobel for sharing the Khare Triton tholin data and Lora Jovanovi{\'c} for sharing the Pluto PAMPRE production rates. 

Data generated as a result of this analysis can be found in the Johns Hopkins University Data Archive \citep{moran2021}.

\textit{Author contributions}: S.E.M., S.M.H, C.H., and N.R.I. conceived the study. C.H. prepared the Triton samples. S.E.M. performed the mass spectrometry measurements with assistance from V.V. and C.W. S.E.M. and M.J.R. performed the spectroscopy measurements. J.S. performed the combustion analysis. 
S.E.M. conducted the mass spectrometry data analysis in consultation with S.M.H. and using software developed by S.M.H. S.E.M. conducted the FTIR analysis with assistance from M.J.R. S.E.M. prepared the manuscript. All authors discussed the results and contributed to editing the manuscript. 


%
%

\bibliography{triton}

%
%
%
%
%

\end{document}